\documentclass[12pt]{iopart}
\usepackage{color,graphicx}
\begin{document}

\title[Size separation in granular matter]{Size separation in vibrated granular matter}

\author{Arshad Kudrolli}

\address{Department of Physics, Clark University, Worcester, MA 01610, USA}

\ead{akudrolli@clarku.edu}

\begin{abstract}

We review recent developments in size separation in vibrated granular materials. Motivated by a need in industry to efficiently handle granular materials and a desire to make fundamental advances in non-equilibrium physics, experimental and theoretical investigations have shown size separation to be a complex phenomena. Large particles in a vibrated granular system invariably rise to the top. However, they may also sink to the bottom, or show other patterns depending on subtle variations in physical conditions. While size ratio is a dominant factor, particle specific properties such as density, inelasticity and friction can play an important role. The nature of the energy input, boundary conditions and interstitial air have been also shown to be significant factors in determining spatial distributions. The presence of convection can enhance mixing or lead to size separation. Experimental techniques including direct visualization and magnetic resonance imaging are being used to investigate these properties. Molecular dynamics and Monte Carlo simulation techniques have been developed to probe size separation. Analytical methods such as kinetic theory are being used to study the interplay between particle size and density in the vibro-fluidized regime, and geometric models have been proposed to describe size separation for deep beds. Besides discussing these studies, we will also review the impact of inelastic collision and friction on the density and velocity distributions to gain a deeper appreciation of the non-equilibrium nature of the system. While a substantial number of studies have been accomplished, considerable work is still required to achieve a firm description of the phenomena. 

\end{abstract}
\submitto{\RPP}
\tableofcontents 
\maketitle

\section{Introduction}

Shake a box of cereal and one invariably sees the largest pieces rise to the top. This spontaneous ordering goes against ones intuition that objects get mixed when jostled in random directions. For binary fluid mixtures (e.g. oil and water), the lighter liquid can be found on the top at equilibrium because of Archimedes' principle. In contrast, the density of the larger pieces of the cereal on top can often have a much greater density than the smaller pieces at the bottom. Cereal is an example of a granular material and size separation is indeed one of the most fascinating examples of self-organization displayed by systems that are out of equilibrium~\cite{jaeger96,shinbrot00,ristow00,rosato00}. 

Like a fluid, granular materials are composed of discrete particles that can move relative to each other. However, the grains interact dissipatively and rapidly come to rest unless mechanical energy is continuously supplied. The dissipated energy raises the temperature of the atoms that constitute the grains, but does not contribute to the properties observed at the granular level. The reason is that thermal energy appears to be negligible compared to the potential energy needed to lift one grain over another in a gravitational field. For example, the thermal energy for a typical granular particle is approximately 18 orders of magnitude less than the mechanical energy. Therefore grains remain locked in place once they come to rest. One consequence is that hydrodynamic descriptions of the stature of the Navier-Stokes equations used to describe fluids do not exist for granular materials. It is interesting to consider that even at rest, a granular system is not at its potential minimum but rather in a meta-stable state\footnote{This gives rise to the possibility that an avalanche can be triggered by thermal expansion due to a temperature change, but we are not after such subtle effects here.}. From these facts one concludes that granular materials are a
system out of equilibrium at almost any level. 

\begin{figure}
\begin{center}
\includegraphics[width=.5\textwidth]{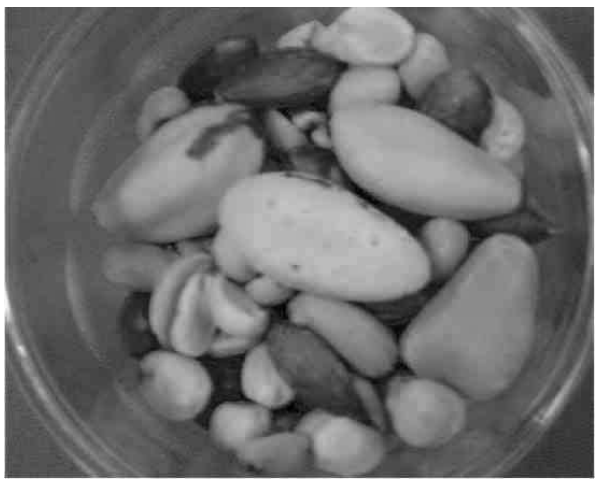}
\end{center}
\caption{The largest nuts rise to the top when a jar of mixed nuts is shaken.}
\label{brazil}
\end{figure}

In typical granular materials, the constituent grains are not identical and may differ in size, density, rigidity, or some other physical property. Such differences can often lead to segregation in granular systems (see Fig.~\ref{brazil}), and size difference is usually considered to be dominant~\cite{williams76,bridgwater93}.  While segregation is often a challenge to avoid, it may be helpful in some situations~\cite{bates97}. For example, separation of chaff from food grains by pouring in a steady wind has been exploited since humans first took to agriculture. Size separation has been extensively investigated in rotating tumblers, the filling and draining of silos, and vibrated granular systems because of their importance in industry~\cite{williams76,bridgwater93,bates97,brown70,cizeau99,samadani99,ottino00}. The phenomena is also important in geophysical processes, for example larger boulders travel further downhill in an avalanche. Considering the numerous contexts in which size separation occurs, understanding the factors which cause and control it is of paramount importance. 

In the context of vibrated systems, one of the earliest report was by Brown on the impact of vibration during transportation of coal pieces~\cite{brown39}. Brown stated that the difference of local packing near a large piece results in more degrees of freedom and thus can cause a large piece to rise even if its density is greater than the small ones. He also noted that if contact forces between particles are eliminated as in some mining processes, then dense particles would sink.  A need for quantitative experiments was expressed to understand the interplay between size and density.

\begin{figure}
\begin{center}
\includegraphics[width=.33\textwidth]{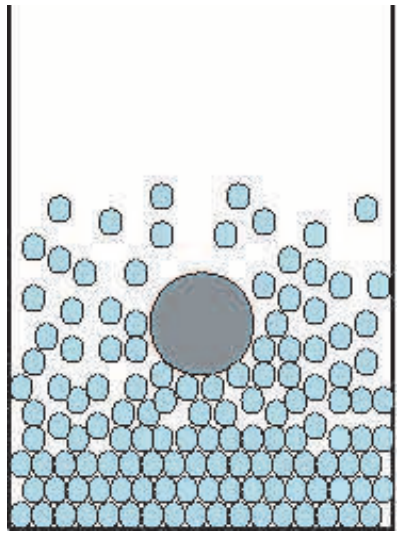}
\end{center}
\caption{The intruder model system investigated by Williams to study size separation contains one large particles in a set of small particles. This system is simpler than a binary mixture with nearly equal volumes of the two kinds of particles because interactions between large particles can be ignored.}
\label{williams_intruder}
\end{figure}

A few decades later, Williams performed systematic experiments and proposed that particle size is the most important property causing segregation~\cite{williams63}. To demonstrate that sieving was not the only reason why large particles are found on top, he performed a model experiment with a single large particle (intruder) and a set of smaller beads inside a rectangular container (see Fig.~\ref{williams_intruder}). When the container was vibrated appropriately, the intruder was always observed to rise upwards. Williams stated that the reason large particles rise is because the small particles below it are stationary and prevent it from falling, while the particles on the side and top move. Thus if the mobile particles cause the large particle to move up, then small particles flow underneath causing the large particle to ascend in a ratchet-like manner~\cite{williams63}.  

While a number of reports on primarily experimental investigations soon appeared~\cite{olsen64,rippie64,ahmad73,parsons76,ratkai76,harwood77,stephens78}, a theoretical study by Rosato, Strandburg, Prinz, and Swendsen clarified the geometrical mechanism for size separation described by previous studies~\cite{rosato87}. They adapted a Monte Carlo simulation technique used in statistical physics to show that a local geometrical void-filling mechanism can lead to size segregation with larger particles on top - the so called Brazil nut effect. The main conclusions of these seminal papers have since become so much part of conventional wisdom, that much surprise and controversy has surrounded any study which seem to show results that may appear otherwise. 

\begin{figure}
\begin{center}
\includegraphics[width=.7\textwidth]{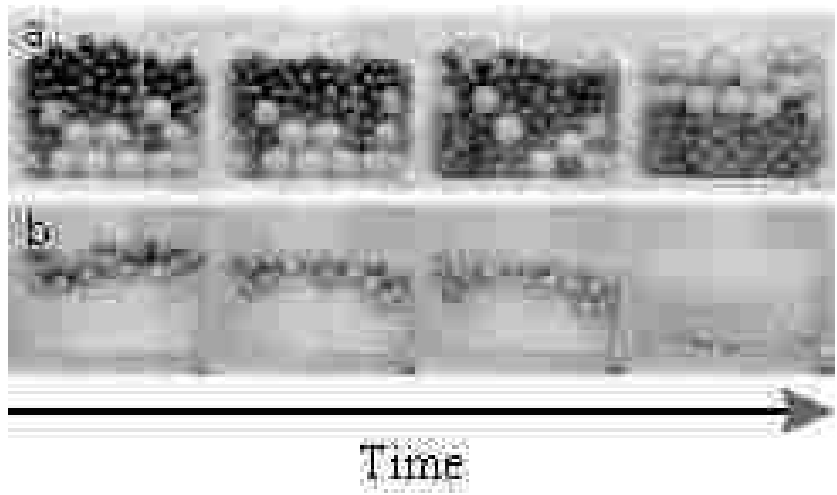}
\end{center}
\caption{Illustration of the complexity of size separation in vibrated system. Large particles can rise to the top or sink to the bottom depending on their density and other physical properties. From Ref.~\cite{breu03}.}
\label{breu_fig1}
\end{figure}

To alert the reader of the subtleties involved, we note some results which appear contradictory. Knight and co-workers~\cite{knight93} observed that boundaries can have a significant impact on segregation. It was shown that frictional interaction of the grains with the side walls can set up convection which can cause an intruder to even go to the bottom, depending on the shape of the container. Based on simulations and statistical physics arguments, Hong and co-workers~\cite{hong01} proposed that a large particle could sink to the bottom provided it was heavy enough - naming it the reverse Brazil nut effect. They derived a relation between particle size and density which would demark the phase diagram. A similar criterion was derived by Jenkins and Yoon~\cite{jenkins02} using kinetic theory for a uniformly heated granular gas in gravity. These theories assume binary collision, ignore friction and may correspond to the regime noted earlier by Brown in which dense particles sink~\cite{brown39}. Recent experiments~\cite{breu03} appear to be consistent with the derived criterion (see Fig.~\ref{breu_fig1}), although it is also clear that several key assumptions in the calculations are not satisfied. Earlier, Shinbrot and Muzzio have shown that a light intruder can go to the bottom of a vibrated deep granular bed displaying what they call a reverse buoyancy~\cite{shinbrot98}. More recently, interstitial fluid including ambient air has been shown to play an important role in determining spatial distributions~\cite{mobius01,yan03}. 

Thus it is clear that factors which appear negligible at first can completely change the observed phenomena. This puts an extremely high burden on both theoretical and experimental studies that seek to compare results while identifying relevant physical parameters.  

At this point, it is also important to note that size separation by itself is neither unique to granular systems nor non-equilibrium systems. For example, colloids consisting of binary mixtures of particles that differ in size exhibit phase separation~\cite{dijkstra99}. In such systems which are neutrally buoyant, gravity does not break the up-down symmetry and consequently lack some of the striking effects seen in granular systems. However, arguments explaining phase separation based on configurational entropy of the system are well developed in equilibrium colloidal systems. Therefore, there is much to be learnt from understanding the effects of dissipation during particle contacts on these ideas. Indeed, mechanisms such as depletion-force induced segregation (first developed to explain phase separation in colloids) are being applied to explain phenomena observed in granular systems~\cite{aumaitre01}. 

To formulate the problem of size separation in vibrated systems, let us consider two species of approximately spherical particles inside a cylindrical container. The species have a mass density $\rho_i$, diameter $d_i$, and number density $n_i$, where $i = L,S$ for the larger and smaller species (the index is dropped if a single species is present). At the grain level, the interaction between particles can be assumed to be a hard-core repulsion interaction. Because of surface roughness, van der Waals forces do not induce any attractive force between particles. Thus the interaction for very rigid particles can be simply modeled as a step function, and by three parameters, the normal and tangential coefficient of restitution $\epsilon_{ij}$, and friction at contact surfaces $\mu_{ij}$~\cite{foerster94}. In principle, dissipative interactions of the particles with the container boundaries can be different from the inter-particle interactions and thus introduce additional variables. 

Starting with the condition where two particle species are made of the same material, one is interested in knowing if the system will size separate, and if the large particles would rise or sink as a function of the size ratio $r_d = d_L/d_S$. Furthermore, the speed of size separation, and how it depends on the strength of vibration and the boundary conditions are of interest. Then one is interested in knowing how additional factors such as the density ratio $r_\rho= \rho_L/\rho_S$ will affect the outcome. Finally, the relevance of additional interactions due to the presence of interstitial fluid, humidity or magnetization on size separation needs to be considered. 

\section{Techniques to investigate size separation}\label{tech}

As discussed in the introduction, granular systems quickly come to rest unless external energy is supplied. In experiments, energy is injected into the system by vibrating the container and many subtleties are involved. The container may be shaken in a vertical and horizontal direction, or even swirled. It is a priori unclear if the nature of the vibration is significant or not. Because most of the experiments have been conducted with vertical vibrations and to avoid unjustified generalizations, we will consider this situation first. We will then consider other types of vibrations and size separation in a later section. 

There are various protocols possible even while vibrating vertically.  At one extreme, the container may be vibrated sinusoidally at high amplitude much greater than the acceleration due to gravity $g$ and frequency ($f >> \sqrt{g/2d_S}$). The system is considered vibro-fluidized in this limit, and particles undergo binary collisions with each other or the container boundaries. The vibro-fluidized limit can always be reached provided the number of particle layers (bed depth) is small enough. The energy input is thought to be fairly continuous, although we will return to this point later. At the other extreme, the container is vibrated with a series of taps at low frequency. Here the particles inside the container gain kinetic energy during the up-stroke, and the introduced energy is quickly lost due to dissipation and particles come to rest before the next tap. In this regime, a wide range of collisions can occur including lasting contacts. A given system can be anywhere in between these two extremes in different spatial regions. In addition, experiments can be conducted by either vibrating the container as a whole, or just the bottom giving rise to various boundary conditions at the side walls. Therefore care should be taken in comparing experiments on size separation with these differences, even when the particle mixtures may be of the same kind. 

The earliest quantitative studies on size separation used manual sorting of the kind of particles as a function of height~\cite{olsen64}. More recently, a number of imaging techniques have been used to measure not only the spatial distributions but also dynamical properties such as rise-time and the rate of progress of segregation. Direct visualization using a CCD-camera in conjunction with quasi-two dimensional geometries or slot-shaped containers has been used to automate measurements~\cite{hsiau97,liffman01}. To investigate three dimensional systems (the insides of which cannot be visualized by light,) radioactive tracer techniques~\cite{harwood77,wildman02} and magnetic resonance imaging (MRI)~\cite{ehrichs95,fukushima99} have been used although not extensively in the context of vibrated systems. 

To mesh the observations together, a physical theory is necessary, but no fundamental theory for even mono-disperse (single sized) granular materials exists~\cite{kadanoff99}. However, in a special regime of high agitation where particles interact by binary collisions, considerable progress has been made in adapting the kinetic theory of molecular gases by incorporating energy loss due to inelastic collisions. At least a few complementary formalisms exist~\cite{jenkins83,haff83,campbell90}, but the underlying assumptions are finally similar to arrive to a point where practical calculations can be made. Dissipation is assumed to be small and particles are assumed to be uncorrelated before and after a collision. In addition, variation in the macroscopic parameters are assumed to be small compared to the grain diameter, and the system is assumed to be smooth. To solve the partial differential equations in space and time, boundary conditions have to be used. Unlike normal fluids, the boundary conditions for granular materials are not well established. Often, periodic or perfectly elastic boundary conditions are assumed, and particles are considered to be either uniformly heated everywhere or at a boundary. 

The kinetic approach has been applied to vibrated granular systems to understand the spatial density and energy variation~\cite{kumaran98}, and more recently to the problem of binary mixtures under gravity~\cite{jenkins02}. It is important to see if these first theoretical descriptions are vindicated by experiments. Furthermore, the domain of validity of the approach needs to be explored. These issues have to be addressed as theoretical progress is made. It is important to stress that the approach is in the regime of binary collisions, and it is unclear if it can be extended to the regime where enduring contact occurs between particles.  

The shear number of parameters involved in doing the simplest experiments with binary systems appears to be beyond the most sophisticated calculations performed to date. For example, the simplest earth bound experiments involve particles of the same material which has low $\epsilon_{ij}$ and $\mu_{ij}$ that differ only in size kept inside an air evacuated container made of the same material which is vibro-fluidized. Particle tracking can be used to find the position and velocity of each particle, but density and temperature in homogeneities exist in principle in this system. Theoretical developments addressing the issue of size separation have up until now~\cite{jenkins02,trujillo02} considered homogeneous distribution for each species, ignored boundaries, and have derived the initial flow of the two species (which is not necessarily the steady state distributions observed in the experiments.) The only way to bridge the gap currently is by using computer simulations, while learning to increase the overlap between experiments and theory. 

Indeed computer simulations using event driven algorithm or discrete element method (also called molecular dynamics) have made substantial contributions in advancing our knowledge of size separation~\cite{poschel95,luding00,hong01,shishodia01}. In the event driven (ED) method, a minimal simulation uses an inelastic hard sphere model to calculate post-collision velocities and then tracks the event when particles will next collide with each other or the wall. This method works primary in the regime of binary collision and low $\epsilon_{ij}$. In the molecular dynamics (MD) simulations, particles are assumed to have finite rigidity and interact for example using the Hertz-Mildrin repulsion force~\cite{johnson85}. In addition, inelasticity and friction are also considered during contact. The equations of motion of the particles are integrated over a small time step to find the resulting position and velocities. The MD method works not only when particles undergo binary collisions but also in a regime of multiple-enduring contacts. For fairly rigid granular particles, the integration time step has to be small which restricts the size and duration of the simulations. However, increasing availability of faster and cheaper parallel machines has widened the scope of such methods. 

In addition to serving as tools to perform numerical experiments in regimes where it is difficult if not impossible to do real experiments (e.g. small $\epsilon_{ij}$, absence of side walls, reduced gravity, etc.) simulations have played an important role in developing and testing not only kinetic/hydrodynamic models, but also models of granular materials where geometry dominates. For example, the Monte Carlo (MC) technique was used to show that local geometric mechanisms can result in size separation~\cite{rosato87,jullien92}. Details of the collision dynamics and energy dissipation are assumed to be not important in obtaining size separation using this technique. 

To go beyond a survey of the literature on size separation, we start with mono-disperse particles, discuss experiments in a few simple limits and the theoretical approaches to understand them. Then we introduce binary systems and with this background go on to discuss size separation. The goal here is to describe key ingredients which may be at work in a given vibrated granular experiment, and no attempt is made to do an exhaustive and historically complete survey. We will then discuss the effect of density, interstitial fluids, humidity, magnetization and electrostatics on size separation. These factors introduce additional forces that can play a significant role in determining the properties of the system. Understanding the effect and strength of each factor is important in arriving at a prediction. 

Self-organization observed in granular systems is far more subtle to understand than classic out of equilibrium phenomena such as Faraday surface waves in oscillated fluids. The primary patterns there can be often understood in terms of a linear instability of an uniform equilibrium state, and much progress in understanding the primary instability has been made~\cite{cross93}. However, even in those cases secondary instabilities have proven to be hard to characterize. The vibro-fluidized limit is a possible limit in which one may be able to attack the general problem of granular flow starting from equilibrium statistical mechanics of dense gases. Therefore, we will pay special attention to this limit in the review. 

\section{Density, temperature and energy flow in vibrated systems}\label{back}

Let us first consider uniform sized grains of diameter $d$ inside a cylindrical container which is being vibrated to understand the effect of dissipation and to get a deeper appreciation of the differences from a normal fluid at equilibrium inside a container. Let us also ignore interstitial air and any phase dependence of the vibration for the moment. (This is not to say that these factors are negligible in realistic systems.)  This serves as a useful starting point because experimental, numerical and theoretical investigations have been conducted for a small number of particle layers that have some overlap. These investigations are predominantly in two dimensional systems, although some work has also been recently performed in three dimensions. In order to compare across various investigations, let us define the number of particle layers $\phi_l$ which is given by the ratio of the total number of particles inside the system to the number of particles required to pack one layer. We also define the driving parameter $\Gamma$ given by the ratio of the peak vibrational acceleration of the system divided by the gravitational acceleration. 

\subsection{Experiments and simulations: Vibro-fluidized limit}\label{dilute}

Warr, Jacques, and Huntley investigated the dynamics of steel disks or spheres confined between glass walls using digital high speed imaging photography and image processing~\cite{warr94,warr95}. Their technique and choice of system allowed them to find not only the position but also the velocity of {\em individual} particles at an unprecedented scale. In experiments conducted at high frequencies, and $\phi_l$ ranging between 0.82 and 2.73, they obtained the packing fraction or density distribution as a function of height, and the velocity distributions in the horizontal $x$-axis and the vertical $y$-axis. Interestingly, a particle density inversion was observed, with a density-peak a few particle diameters away from the driving wall before smoothly decreasing to zero with height~\cite{warr95}. Therefore the density distributions are neither uniform, nor can be fit by an exponential. 

\begin{figure}
\begin{center}
\includegraphics[width=.95\textwidth]{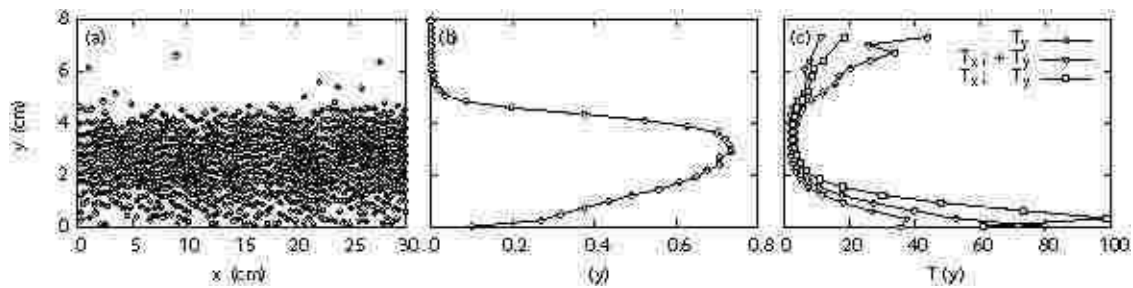}
\end{center}
\caption{(a) Positions of spheres constrained to two dimensions using high speed imaging. (b) The number density as a function of height of a granular gas shows a peak. (c) The granular temperature is observed to be anisotropic and shows an temperature inversion above the peak in the density. Adapted from Ref.~\cite{blair01}.}
\label{blair_fig1}
\end{figure}

The velocity distributions in both horizontal and vertical directions were claimed to be fitted by Gaussian although poor statistics make it impossible to detect any systematic deviations. Many experiments since then, using the same technique with improvements in resolution and statistics, have shown the distribution to be in fact locally non-Gaussian for a wider range of $\phi_l$, and driving conditions~\cite{losert99,olafsen99,rouyer00,blair01}.  Using the width of the velocity distributions, it is possible to define a temperature~\cite{ogawa78}. This temperature called the ``granular temperature" (to distinguish it from thermal temperature) is given by $T_{x,y} = \frac{1}{2}m \langle v_{x,y}^2\rangle$, where $m$ is the mass of the grain (the Boltzmann factor is set to 1). Using such a definition and fitting their measured distributions, Warr, {\em et al.}~\cite{warr95} showed that $T_x$ and $T_y$ are not equal. Later experiments have shown that even the vertical distribution is anisotropic (see Fig.~\ref{blair_fig1}). Thus temperature is not a scalar in vibrated granular systems, and equipartition of energy is not observed.  Indeed none should be expected, unless of course one persists in viewing it from the cherished point of thermal systems at equilibrium. 

The reason for the anisotropy in the granular temperature arises due to the inelasticity of the particles, and the nature of the driving. Even the least inelastic particles (such as the Chrome Steel particles used by Warr, {\em et al.}) have $\epsilon \approx 0.93$. Furthermore, energy is injected into the particles at the bottom (ignoring the side walls). In the experiments of Warr, {\em et al.} and its variations which are relevant to the discussion of size separation, the bottom wall oscillates in the vertical direction, imparting an overall positive velocity to the particles. (Even considering friction,  the velocity of the horizontal direction does not change significantly.) Especially in systems with $\phi_l$ relevant for size separation, the particles change directions (to horizontal and downward) after suffering an inelastic collision. Thus energy is fed into the other directions via collisions which always on an average take energy away. Thus the energy is lower in the other directions by a factor depending on inelasticity and the height away from the boundary where energy is injected. Such effects in principle do not pose a problem in the development of a theory and, as we will discuss later, is already underway. Also because the anisotropy arises due to the nature of the energy injection, the direction of vibration can be anticipated to have an effect on diffusion and size separation. 

Qualitatively similar behavior has been observed in simulations~\cite{luding94,luding95}, which also considered the effect of adding or removing surface friction and the effect of the dissipation at the boundaries~\cite{mcnamara98}. Simulations and experiments also showed that the granular temperature as a function of height, has a minima above the point where the density peaks~\cite{helal97,baldassarri01,blair03}. Such a temperature inversion above the peak in the density is a consequence of the dissipation, and may be described by modifying the Fourier's law for heat flow in an elastic hydrodynamic systems~\cite{ramirez03}.

\subsection{Kinetic theory: Single species}\label{kinetic}

As mentioned in Section~\ref{tech}, considerable theoretical progress has been made in a regime where granular particles interact by inelastic binary collision, as is the case in the experiments and simulations discussed above. Methods of kinetic theory for dense fluids have been modified to incorporate the loss of energy during collisions. The starting point is the steady state Boltzmann equation which assumes ergodicity and describes the evolution of the particle velocity distribution function $f({\bf x}, {\bf u})$~\cite{kumaran98}:
\begin{equation}
\frac{\partial(u_i n f)}{\partial x_i} + \frac{\partial(a_i n f)}{\partial u_i} = \frac{\partial_c(n f)}{\partial t},
\end{equation}
where, $n$ is the particle number density function, ${\bf u}$ and ${\bf a}$ are the velocity and acceleration of the particle respectively. The index $i$ denotes the component of the vectors, and repeated index refers to a dot product. This equation relates the convective transport of the particles in real space, the velocity transport in velocity space due to acceleration, and the rate of change of the distribution due to collisions. The rate of change term considering binary collisions can be in general written as a function of the particle density, the equilibrium radial distribution function at contact, the integral of velocity of the second particle, and the relative orientation of the particles. 

At this point assumptions have to be made with respect to the particle pair distribution functions in order to proceed with calculations. Therefore the complete pair distribution functions at collisions are written as the single particle velocity distribution function for each particle and factors that take in to account the volume excluded by the particles. The excluded volume effects can be significant at the high density relevant for granular systems. It is assumed that even neighboring particles are uncorrelated which may not be accurate especially at high densities. However, as we will discuss later, there appears to be some work which supports this approximation at low densities. 

The next simplifying step is to assume a Maxwellian distribution function for the single particle distribution function:
\begin{equation}
f({\bf u}, {\bf x}, t) = \frac{n m}{2 \pi T} {\rm exp}\frac{- m \langle{\bf u}^2\rangle}{2 T},
\end{equation}
where $T$ is the granular temperature. Making further approximations for the excluded volume effects for a hard sphere, one then arrives at the constitutive relations for the rate of dissipation, pressure tensor, etc. 

Kumaran~\cite{kumaran98} has calculated the correction to the density, temperature and the moments of the distribution by perturbing around the Boltzmann distribution for the density, and the Maxwell-Boltzmann distribution for the velocity distribution. He keeps the first corrections due to dissipation and appears to show trends in qualitative agreement with observations in Ref.~\cite{warr95} and Fig.~\ref{blair_fig1}. Given the number of untested assumptions in the calculations that we sketched briefly, even the qualitative agreement is remarkable. 

However, a number of experiments and simulations have since shown that the velocity distributions are non-Gaussian for inelastic particles even in small regions where density and velocity inhomogeneities are apparently absent~\cite{rouyer00,blair01}. These observations may imply correlations and thus a breakdown of the simplifying assumptions used in such calculations.

Using a direct simulation Monte Carlo technique, Baldassarri, {\em et al.}~\cite{baldassarri01} have obtained density distributions and velocity distributions similar to those observed in vibro-fluidized experiments~\cite{blair01}. The simulation technique breaks the motion of the particles into two parts. The first corresponds to free flow under gravity. The second part corresponds to collisions, where a particle is assumed to collide with a second particle in its neighborhood with a probability which depends on its size. The post-collision velocities are computed assuming a random collision angle and the normal coefficient of restitution. It is interesting to note that {\em molecular chaos} is assumed in the simulations and yet non-Gaussian distributions and long range velocity correlations~\cite{puglisi_pc} similar to those in the experiments are observed. When molecular chaos can be assumed, it has been shown that the solutions of the Monte Carlo simulations converge to that of the Boltzmann equation. Therefore the study appears to support the approximation which is usually made in calculations starting with the Boltzmann equation where the two particle collision function is replaced by a multiplication of single particle functions. However, much work is still required to clarify why the velocity distributions show correlations. 

In dense systems it is not possible to homogeneously vibro-fluidize particles even via violent shaking, and binary collisions cannot be assumed. Furthermore, for deep beds, it is possible that particles remain in contact through a significant fraction of a vibration cycle even at high frequencies. Thus the kinetic approaches discussed may not be very effective in dealing with such a situation. Currently it is not even clear where the transition from the binary limit occurs as the number of particle layers (bed thickness) is increased.

\subsection{Boundary and interstitial air driven convection}\label{convection}

Up until now, we have ignored the effect of boundaries on density and velocity distributions. However, it is becoming increasingly clear that side walls do play a role in setting up convection currents especially in dense systems. The fact that convection occurs in granular matter is known from Faraday's first experiments on powders~\cite{faraday31}. Convection can lead to heaping where the top surface is inclined with the horizontal. In a number of systems, convection enhances mixing which counteracts segregation. As we will discuss in later sections, convection may either cause or diminish segregation in vibrated systems. 

Molecular dynamics simulations of disks inside a two dimensional container have shown that shear friction between particles during collisions can lead to convection~\cite{gallas92,taguchi92}. The simulations performed with $n_l \approx 13$ and $7$ showed a downward motion existed at the fixed vertical side walls, and an upward motion near the center. The observations were explained in terms of differences in the shear friction encountered by particles during the up and down part of the vibration cycle. The hypotheses was that the friction would be stronger on the upward cycle because particles are more compressed against each other and the side walls. On the other hand the friction may be smaller on the downward cycle because particles are much more loosely packed and therefore the shear friction is much less. 

\begin{figure}
\begin{center}
\includegraphics[width=.7\textwidth]{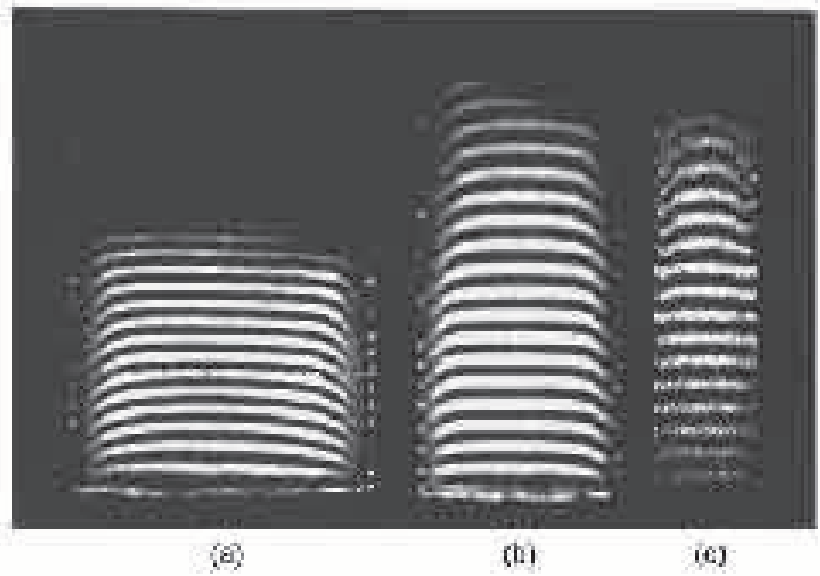}
\end{center}
\caption{Convection rolls develop when a dense bed of granular materials is vibrated vertically. MRI techniques reveal the flow to be downwards near the side walls and upward in the center. From Ref.~\cite{knight96}. }
\label{knight_fig1}
\end{figure}

Using MRI~\cite{ehrichs95,knight96}, it has been shown directly that a thin boundary layer develops where the flow is downward when a cylindrical container filled with poppy seeds is vibrated at low frequencies. (The appearance of localized boundaries between moving and static regions is indeed one of the hallmarks of granular systems~\cite{nedderman92}.) These experiments are performed in some sense in the opposite limit than the vibro-fluidized cases discussed in the previous sections, with a layer of the same kind of particle glued to the side walls of the container to obtain a rough boundary condition. The investigation showed that the upward flow in the center decayed exponentially with depth (see Fig.~\ref{knight_fig1}).

In addition, it was also noted that the shape of the container has a significant impact on the strength and direction of the convection roll~\cite{knight96}. In particular, when the side walls were inclined away from the vertical (for example in a V-shaped container,) grains would {\em rise} near the side walls and fall in the center, a so-called reversed roll in contrast with a normal roll discussed earlier. The effects of the container boundary was examined in detail using computer simulations by Grossman~\cite{grossman97}. The numerical and the experimental data for the mean velocity as a function of distance from the center could be fitted with a parabolic function for low container diameters, and by a hyperbolic cosine function for wider containers~\cite{knight96,grossman97}. 

Grossman found that particle density variation within the granular bed is responsible for convection. Furthermore, she identified that the physical mechanism at work was the density dependent wall friction for both normal as well as reversed rolls. The direction of the roll depended on when the peak in density occurred in relation to the relative velocity of the side walls, and was influenced in turn by the angle of inclination of the side walls and its roughness~\cite{grossman97}. In the vibro-fluidized regime discussed in Section~\ref{dilute}, there is very little density variation with the drive cycle, and therefore convection should be absent. These ideas appear to be consistent with at least some observations~\cite{knight95,blair03}.

In addition to friction with side walls, interstitial air can also play an important role in convection. In experiments conducted with a vibrated annular container, Pak, {\em et al.}~\cite{pak95} observed that heaping diminished considerably and even disappeared when the interstitial air was allowed to escape by using permeable walls or by evacuating the air out of the container. It was found that particles less than 1mm in size were especially susceptible to interstitial air effects, and the effect was increasingly important as bed depth increased. They concluded that the reason for the convection was not Stokes drag (which is important for example in gas fluidized beds) but rather the pressure difference that arises due to trapped air between the grains and the boundary. Thus a large excursion of the vibrating boundary which is the case at low frequencies can enhance the effect of even ambient air.

It is also possible that convection can occur in an essentially infinitely wide container with particles interacting dissipatively. As discussed in previous sections, a density inversion exists near the bottom wall. Therefore it is possible that buoyancy driven convection may occur. Such an interesting possibility may explain some numerical simulations results~\cite{ramirez90} and has been considered theoretically by Khain and Meerson~\cite{khain03}.

\subsection{Compaction}\label{compaction}

In the regime where vibration frequency is small, the system does not remain fluidized throughout the cycle even for large amplitudes. The shaking is assumed to promote local rearrangement and geometry plays a dominant role in determining the properties of vibrated granular materials. This is the regime in which MC simulations were applied to size separation~\cite{rosato87}. But first let us consider the packing and density variation for a set of single sized particles in a vibrated container. For a normal fluid, the volume and density are well defined quantities. However, as can be demonstrated by filling a tall narrow container with grains, the total volume occupied can drop by several percent after tapping the container a few times. 

In fact, careful experiments~\cite{knight95} have revealed that the average density can increase over a long series of taps and reach a steady state which depends on $\Gamma$. Large fluctuations are observed even about the steady state value which is around 0.63~\cite{nowak98}. The packing is notably lower than for regular hexagonal packing of uniform spheres. (The grains are considered to be randomly packed although it is a density which is not precisely defined~\cite{torquato00}.) The experiments also showed that the local densities varies considerably, although the average density was more or less uniform. In the context of this review, it is interesting to note that a geometrical theory relying on free volume arguments was able to describe the kinetics of the density~\cite{ben-naim98}.  Noting that a rearrangement of a growing number of particles is necessary to accommodate  additional particles, Ben-Naim, {\em et al.}~\cite{ben-naim98} were able to explain the form of the density change in the experiments. Thus it appears that models that neglect details of the kinetics can play an important role in describing properties of granular materials.

Such an approach has in fact been suggested by Edwards and coworkers~\cite{edwards93,mehta90} for granular materials undergoing slow motion.  In this approach it is assumed that the role of vibration is to introduce noise which leads to collective rearrangement of particles. They introduced a compactivity parameter which plays the role of a temperature which is said to physically capture the state of the system better than the thermal or the granular temperature introduced earlier. The temperature $T_X$ (sometimes referred to as Edwards temperature) is defined using the different possible ways of arranging the grains in the system (configuration statistics). This general formalism has been also used to describe segregation in binary and polydisperse granular mixtures~\cite{shapiro96}. In many ways there is much in common with the approach developed by Savage and Lun~\cite{savage88} to quantify the void-filling segregation mechanism at work in inclined chute flows. However, predictions that are testable experimentally need to come out of the model, and it is possible that size separation in dense granular mixtures is the system to develop these ideas further. 

\subsection{Granular temperature in binary mixtures}\label{binarytemp}

In an effort to define global variables, we introduced granular temperature $T_g$ in terms of the mass and the velocity of the particles as being more relevant (rather than the temperature of the constitutive atoms) to describing the state of a vibro-fluidized granular system. Earlier we saw that the granular temperature is anisotropic even in the simplest vertically  vibrated system because of inelasticity. Therefore the question arises, is the $T_g$ the same or not for all particles in a binary mixture that may differ by size, density or inelasticity as we will encounter in size separation?

Consider a vibrated system consisting of uniform particles and a single particle with properties different than the rest. Losert, {\em et al.}~\cite{losert99} investigated such a situation using experiments in which a few steel or glass spheres of various sizes were vibrated in a container along with a sub-monolayer of glass particles. Under the applied vibration conditions, the particles were confined to essentially two-dimensions at the bottom of the container thus allowing the all particles to be visualized directly using a CCD camera. The velocities of the two species was obtained by tracking the position of individual particles over many time intervals between collisions, enabling the instantaneous velocities to be determined. The temperature was thus determined by examining the velocity distributions. It was found that {\em the temperatures of the two kinds of particles were different}. Furthermore, it was found that the ratio of the temperature of the heavier particle to lighter particle increased with the mass ratio. Thus equipartition was not observed in these experiments. 

\begin{figure}
\begin{center}
(a)\includegraphics[width=.45\textwidth]{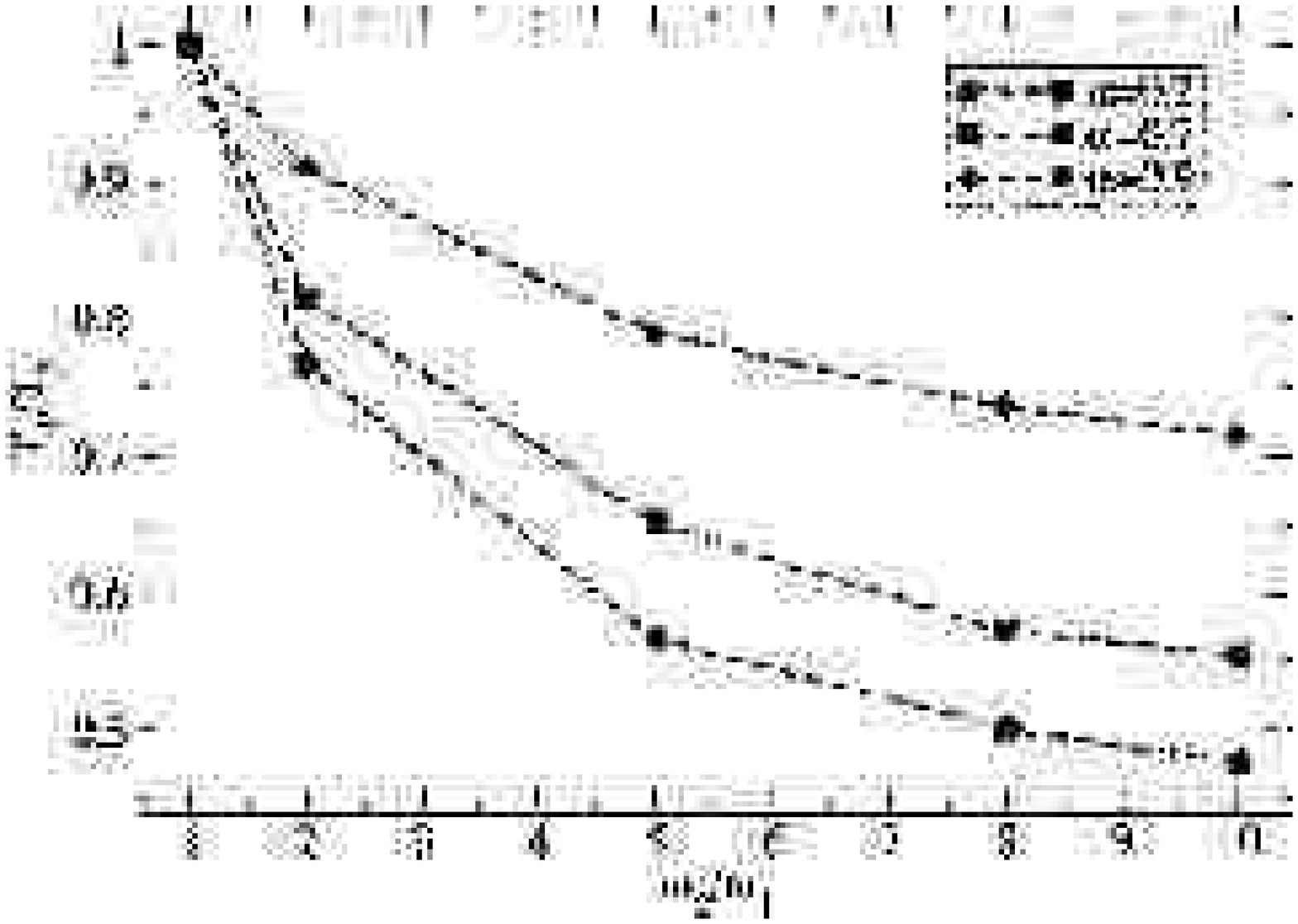}
(b)\includegraphics[width=.45\textwidth]{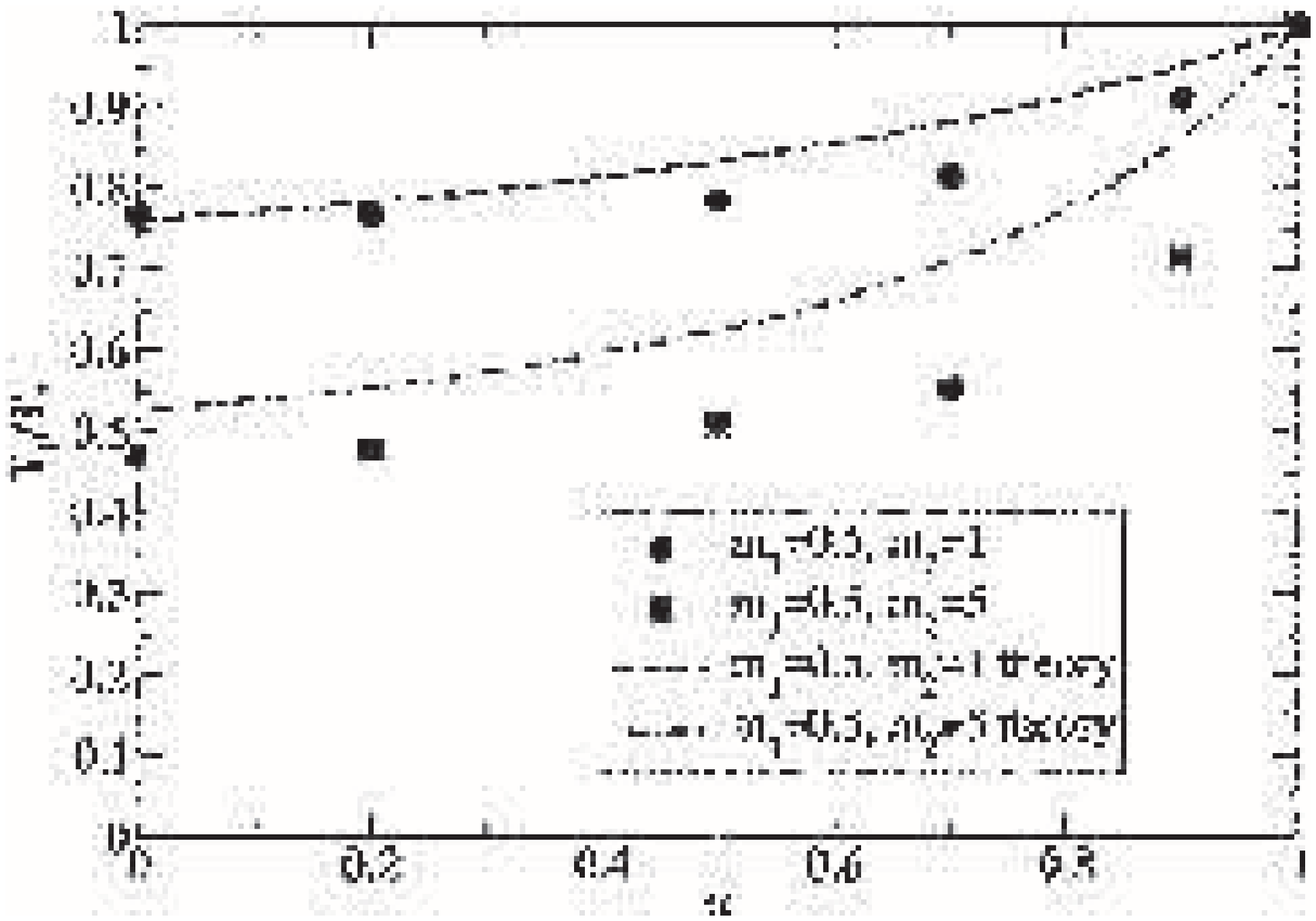}
\end{center}
\caption{(a) $T_L/T_S$ as function of mass ratio for binary mixtures obtained using DSMC simulations by Pagnani, {\em et al.}.~\cite{pagnani02}. (b) $T_L/T_S$ versus restitution coefficient $\alpha$ for binary mixtures obtained using DSMC simulations by Pagnani, {\em et al.}.~\cite{pagnani02}. }
\end{figure}

Similar effects were observed to persist in the bulk of a two dimensional vibro-fluidized binary mixture by Feitosa and Menon~\cite{feitosa02}. They also observed that the temperature ratio was relatively insensitive to the inelasticity of the particles, the vibration velocity, and the number fraction of each component, and only depended on mass ratio. Using positron emission tomography, Wildman and Parker~\cite{wildman02} have also found that the temperature of the large particle was consistently higher than that of the lighter particle. In contrast with the previous mentioned measurements in a 2D system, they found that number density did have an impact on the measured temperature. They argued that the nature of packing may have an impact on the granular temperature. 

Some of these issues at least in low densities, have been clarified by Pagnami, {\em et al.}~\cite{pagnani02} using a direct simulation Monte Carlo (DSMC) technique discussed earlier in the review. For a binary mixture in which energy is supplied by a homogeneous heat bath, they found that the large particles could either have higher or lower energy than the smaller particles depending on the model they adopted for injection of energy. In particular, when the friction and the energy supplied was proportional to the mass, they found $T_L/T_S$ to be greater than one. On the other hand if both species were assumed to receive the same energy, and because a heavier particle dissipates more energy, it would have a lower temperature. 

Noting again that DSMC technique yields the same results as the situation of the Boltzmann equation for averaged quantities, they derived analytical estimates for mass ratio dependence as well as inelasticity~\cite{pagnani02}. Their results matched published experimental results where available. However they also found dependence of $T_L/T_S$ on $\epsilon$ contrary to conclusions based on experimental data~\cite{feitosa02}. It is possible that a wide enough range in $\epsilon$ has not been explored in experiments to uncover the dependence. Qualitatively similar behavior was observed with DSMC when energy was supplied by a vibrated bottom boundary in the presence of gravity. Differences between the densities of the two species as a function of height developed, but $T_L/T_S$ was constant and always greater than one. 

Thus we conclude that the granular temperature of the two species of particles in a realistic system have to be assumed to be different unless the differences can be shown to be either  negligible or inconsequential. 

\section{The intruder model system}

\begin{figure}
\begin{center}
\resizebox{7 cm}{5 cm}{\includegraphics{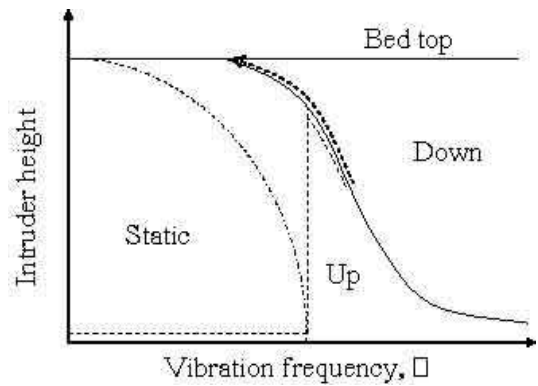}}
\end{center}
\caption{The motion of a large particle in a vibrated bed of small particles observed by Williams~\cite{williams63}. Williams found that the large particle rises above a critical amplitude and reaches an equilibrium height which depends on the vibration strength. A sketch of the particle positions is shown in Fig.~\ref{williams_intruder}.}
\label{williams_fig2}
\end{figure}

With this background we go on to discuss size separation in vibrated granular matter using the intruder model system where few if not only one large particle is present. Williams~\cite{williams63} has stated that when a large particle is placed at the base of a vibrated container,  the particle will always rise and reach a height in the bed that depends on vibration strength (see Fig.~\ref{williams_fig2}). However, as quantities are not precisely defined, it is difficult to get more than a qualitative picture because it is unclear in which limit the experiments were done. 

\subsection{Vibro-fluidized case}

\begin{figure}
\begin{center}
(a)\includegraphics[width=.45\textwidth]{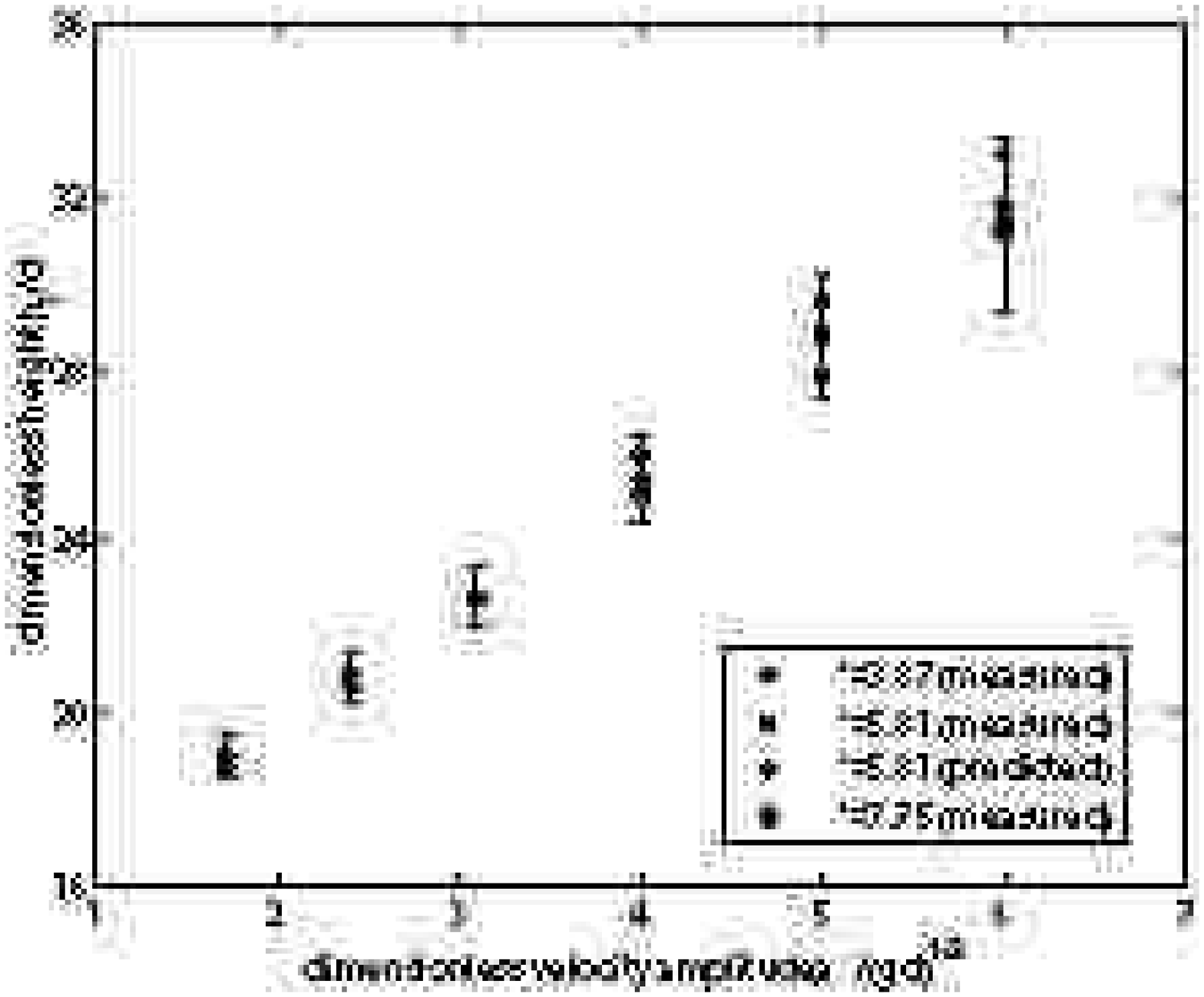}
(b)\includegraphics[width=.45\textwidth]{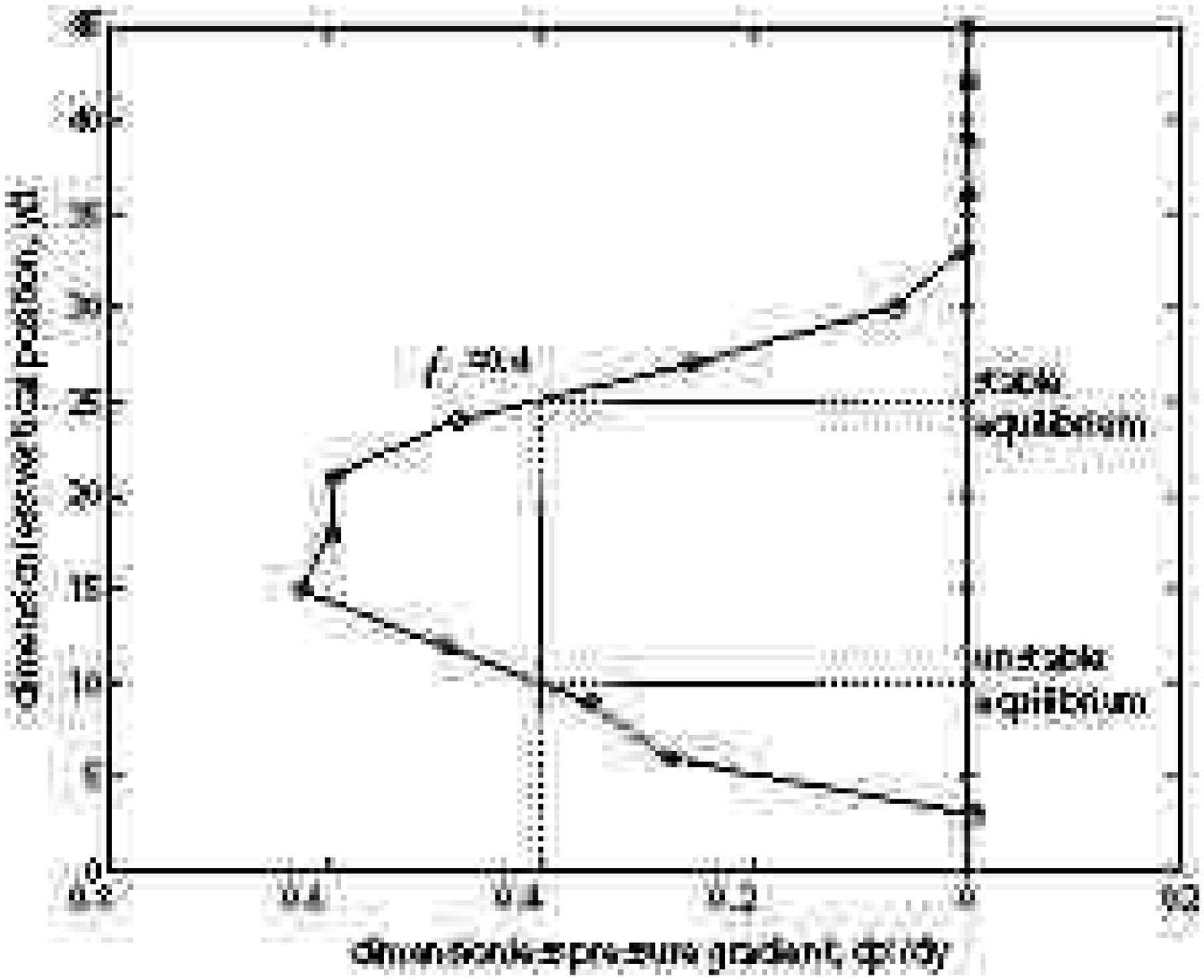}
\end{center}
\caption{(a) The equilibrium height of the large particle as a function of dimensionless velocity amplitude ($m_L/m_S = 10$, $r_d = 5$). $r_\rho <1$ in these simulations. The data for various driving frequencies are shown. (b) The equilibrium height of the large particle as a function of pressure gradient. From Ref.~\cite{shishodia01}.}
\label{buoyant}
\end{figure}

Let us first consider the dilute vibro-fluidized regime where boundary effects can be ignored. As discussed in Section~\ref{back}, when $n_l$ is increased such that $n_l\,(1 - \epsilon_{L,S})\,<\,1$, then the system may be vibro-fluidized for high $\Gamma$. Using MD simulations (or discrete element computer simulations), Shishodia and Wassgren~\cite{shishodia01} studied vibro-fluidized particles inside a square box with $n_l \approx 15$, and $\epsilon_{S,L} = 0.95$. The large particle, after an initial transient, is observed to rise approximately linearly to an equilibrium height and fluctuate about that point due to oscillations of the container and the impact of the surrounding particles. The equilibrium height was also observed to change monotonically with vibration amplitude [see Fig.~\ref{buoyant}(a)], and could be explained by considering a model where the impurity weight was balanced by a buoyant force due to the collisions with the surrounding particles, and its stability around that point. Therefore 
\begin{equation}
\frac{dP}{dy} = - \rho_L g,
\label{pressure}
\end{equation}
where, the pressure $P$ is computed by summing the streaming normal stress and the collisional normal stress. Next they obtained the pressure for the small particles using the MD simulations (neglecting the presence of the large particle), and showed that equilibrium position was indeed predicted by their model [see Fig.~\ref{buoyant}(b)]. Interpreted in terms of this model, the location of the large particles is determined by the unusual pressure gradients in a vibro-fluidized system. However, it is to be noted that this analysis was conducted with $r_\rho < 1$, where a normal fluid would also rise to the top at equilibrium. Ohtsuki, {\em et al.}~\cite{ohtsuki95} have conducted simulations for $0.5 < r_\rho < 4$ and $r_d = 1, 1.5, {\rm and}\, 2$. They found that the intruder sinks deeper into the bottom half of the container as $r_\rho$ is increased. Thus at least for the dilute vibro-fluidized regime, the simulations suggest that the average height of a large particle is a smooth function of both size and density ratios of the particles.

Following the discussion in Section~\ref{binarytemp} it appears that the presence of a large particle will affect the analysis somewhat, as it will most likely have a higher granular temperature. These effects may have to be incorporated into the model before comparing with experiments. However, there have been no direct experiments reported on this regime although such experiments are feasible. For example, pressure gradients for the mono-disperse case have been measured by direct particle tracking techniques ~\cite{blair03}, and appear to have the form seen in Fig.~\ref{buoyant}. 

In principal, the gradient of the pressure in Eq.~\ref{pressure} should be also calculable from the kinetic approach described in Section~\ref{kinetic}. This would be a good test for feasibility of the kinetic approach. However, such a study has not been accomplished as yet although there are some reports towards achieving this goal in binary mixtures. We will return to this point after discussing size separation in dense gases. 


\subsection{Dense systems}

\begin{figure}
\begin{center}
\includegraphics[width=.7\textwidth]{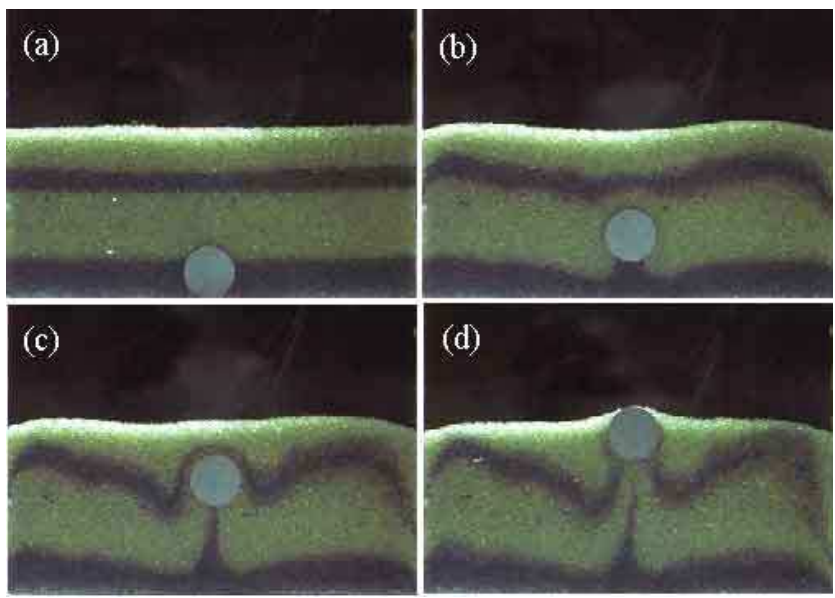}
\end{center}
\caption{A time sequence of the rise of a large disk in quasi-two dimensional experiments reported by Liffman, {\em et al.}~\cite{liffman01}. From the images it is clear that while some convection exists, the large particle essentially moves through the smaller particles as it rises to the top.}
\label{liff_image}
\end{figure}

Now let us examine the issue of size separation in the dense regime ignoring the effects of interstitial fluid for the present. This is a regime in which $n_l (1 - \epsilon) > 1$ and the particles cannot be assumed to undergo binary collisions at all times and locations. 

\subsubsection{Experiments in the non-convective regime}\label{seg_non_con} 

Figure~\ref{liff_image} shows the rise of a large steel disk in a bed of small glass particles in experiments performed by Liffman, {\em et al.}~\cite{liffman01}. The images capture the motion of the large particle through the small particles, and is a strikingly visual demonstration of the Brazil-nut phenomena which has been described by many others going back to Brown~\cite{brown39}. While counterflow to preserve volume and some downward motion of small particles near the side walls are visible, such effects appear to be not very important in determining the upward motion of the disk. Thus to simplify the analysis, let us first focus on experiments in this regime where  convection is either negligible or absent. 

\begin{figure}
\begin{center}
(a)\resizebox{7 cm}{5 cm}{\includegraphics{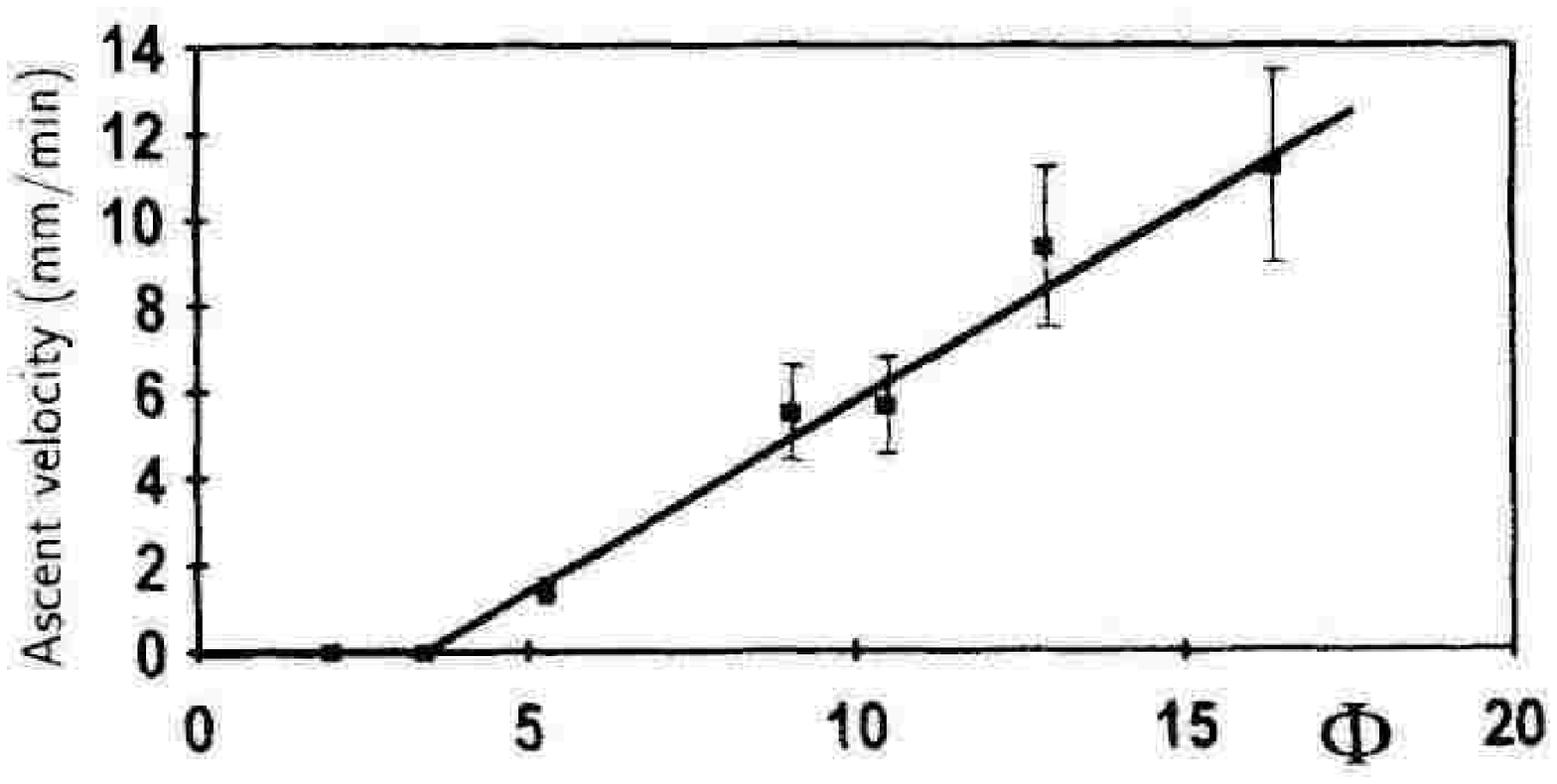}}\vspace{6 cm}
(b)\resizebox{7 cm}{5 cm}{\includegraphics{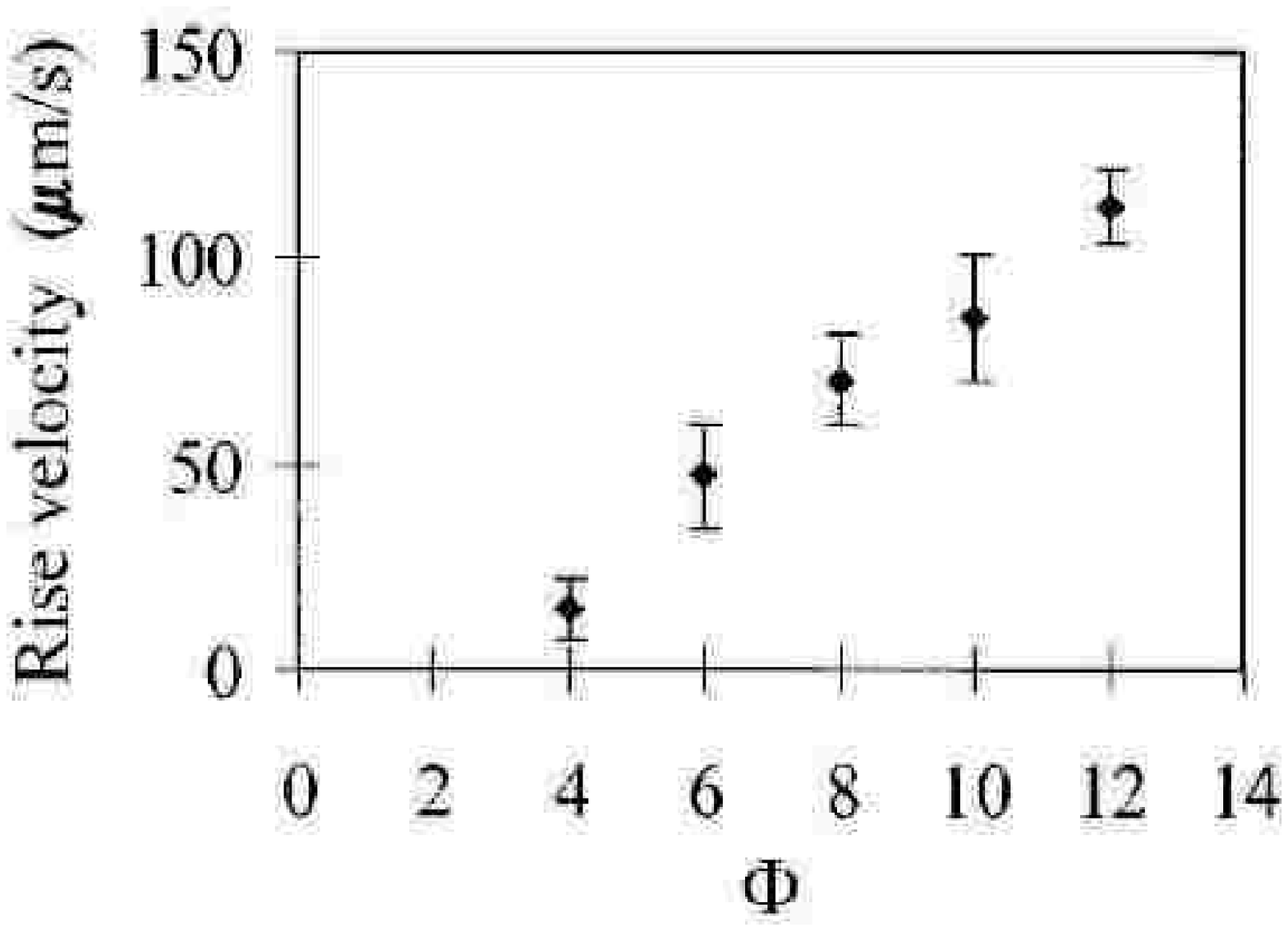}}\vspace{6 cm}
\end{center}
\caption{(a) Rise-velocity of an aluminum disk as a function of diameter ratio $\Phi$ measured in quasi-two dimensional experiments~\cite{duran94}. (b) Rise-velocity of an acrylic sphere measured inside a cylinder~\cite{vanel97}. The rise-time is observed to decrease with size ratio.}
\label{rise_time}
\end{figure}

Let us also further examine the case where the particles are made of similar material and density. Such experiments have been performed in two dimensions by Duran, {\em et al.}~\cite{duran94}, and in three dimensions by Vanel, {\em et al.}~\cite{vanel97}. In these experiments, a ``non-convective" regime was found for appropriate driving frequencies and amplitudes. The non-convective regime appears to correspond to high $f$ and low $A$. A size dependent rise-time/velocity is observed in both cases (see Fig.~\ref{rise_time}). The small particles are seen to be regular-packed with defects around the large particles. 

\subsubsection{Experiments in the convective regime}\label{seg_con}

\begin{figure}
\begin{center}
(a)\includegraphics[width=.6\textwidth]{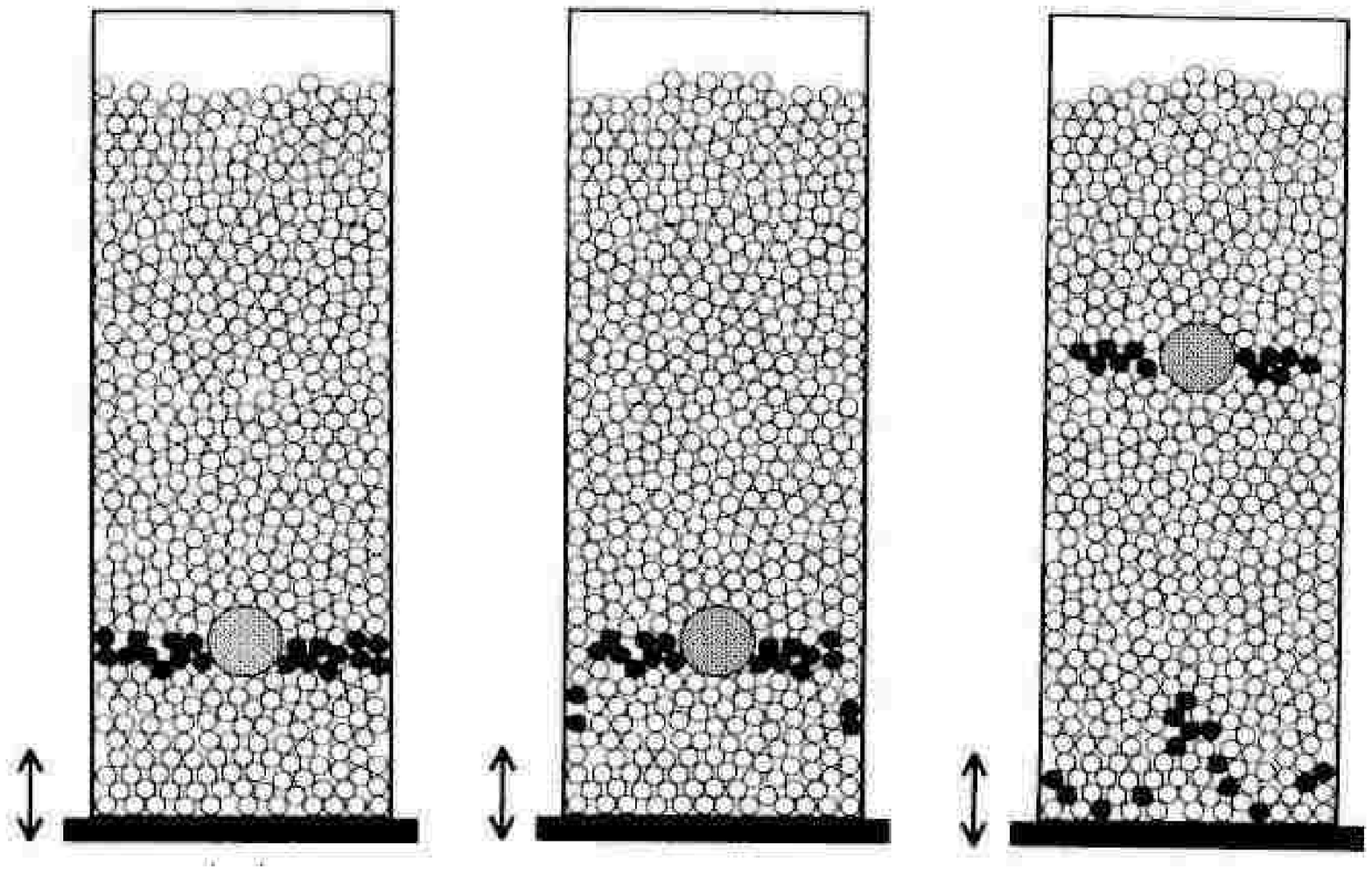}
(b)\includegraphics[width=.25\textwidth]{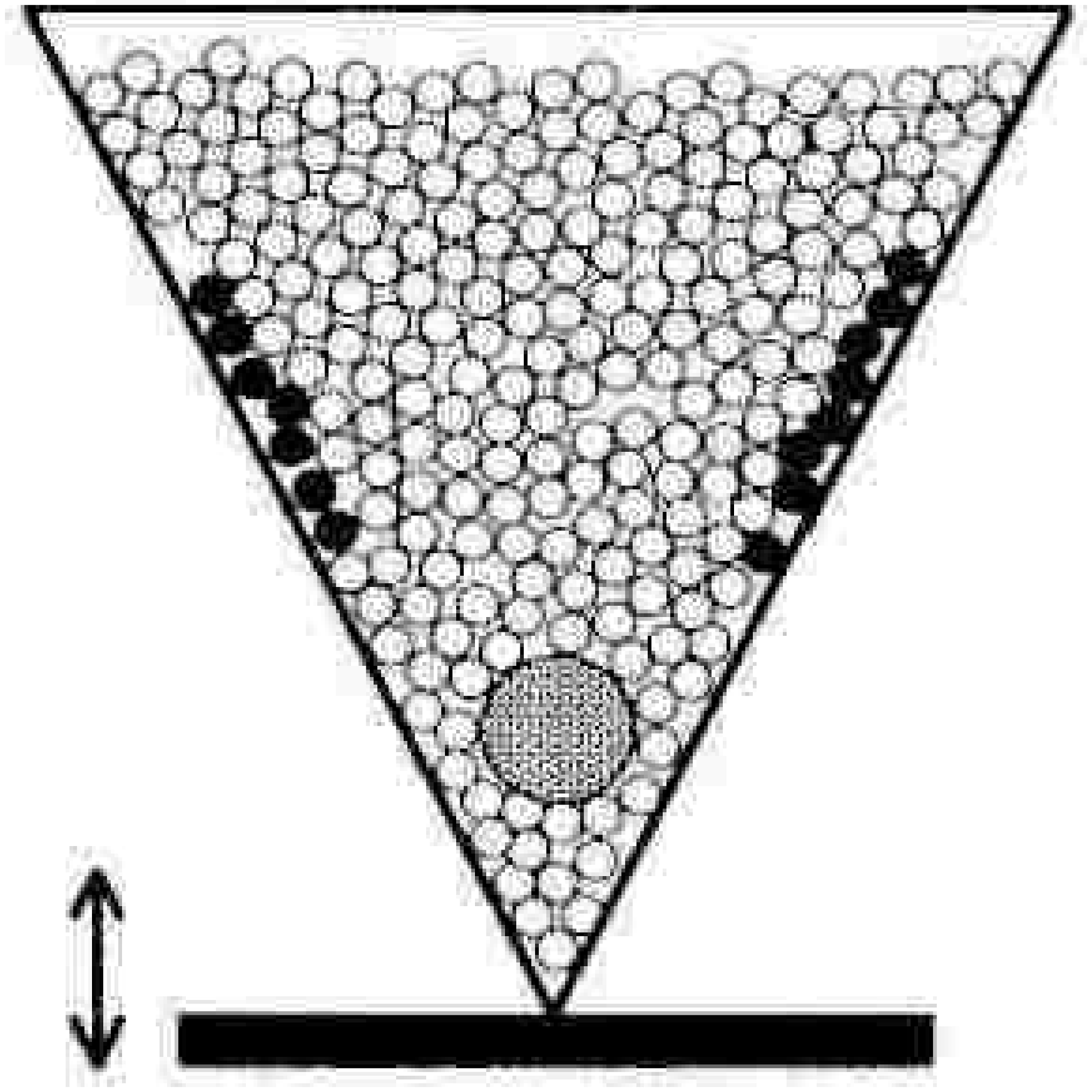}
\end{center}
\caption{(a) Rise of a large particle due to convection. Small particles fall in a thin boundary layer and rise at the center of the container. (b) The large particle can sink to the bottom of a $V$-shaped container when reverse convection is setup. From Ref.~\cite{knight93}.}
\label{conv_seg}
\end{figure}

Now let us discuss the situation when convection is present. In vibration experiments performed with glass beads inside a cylinder, Knight, {\em et al.}~\cite{knight93} observed that the presence of convection had a strong influence on the location of the large particle. The cylinder was vibrated (tapped) with a high amplitude and low frequency. They found that an intruder would rise to the surface almost preserving its relative position with the surrounding small particles until it moved to the top and then to the side of the container where it would remain [Fig.~\ref{conv_seg}(a-c)]. As discussed in Section~\ref{convection}, frictional interaction with the side wall plays an important role in setting up convection currents. Moreover, the shape of the container has an important impact on the direction of the flow near the wall. In the situation shown in Fig.~\ref{conv_seg}(a-c), the vertical walls promoted convection rolls in which particles fall in a thin layer next to the boundary, and therefore due to conservation of volume have to rise at the center. From their observation Knight and coworkers concluded that the intruder rose to the top due to convection and was trapped there because it could not fall through the thin boundary flow. On the other hand when a reverse roll was set up by using a container with outward sloping side walls, the intruder was observed to move to the bottom and get trapped there because it was too large to fit in the upward flowing thin boundary layer. 

If particle motion is only due to convection, the rise of the large particle inside a cylindrical container would depend then only on the strength of the convection and independent of the size of the particle. Indeed, Knight, {\em et al.}~\cite{knight93} observed that the rise time was be independent of the size of the particle. They further observed that when convection was absent (by decreasing friction at the side walls), the particles did not move relative to each other and remained trapped in their original positions. 

Detailed experiments later by Cooke, {\em et al.}~\cite{cooke96} in model two dimensional systems also appear to show convection-like motion for high as well as low frequency of vibration. In this case, a large intruder particle was observed to rise with velocities similar to surrounding small particles. They did not find evidence for the intruder motion through the small particles via local grain rearrangement. Experiments by Duran, {\em et al.}~\cite{duran94} and Vanel, {\em et al.}~\cite{vanel97} have found convective as well as non-convective regimes where rise-times were dominated by mean-flow. Poschel and Herrmann~\cite{poschel95} performed molecular dynamics simulations and found that convection can be triggered by the motion of the large particle. 

While it is clear that size separation can occur without convection, as is illustrated in Section~\ref{seg_non_con}, the presence of convection can certainly alter the velocity of the large particles. We will return to the importance of convection on size separation after discussing convection in binary mixtures where the interactions between large particles is important as well. 

\subsection{Geometric models of intruder motion}

Next we discuss the possible mechanisms by which the large particle moves relative to the small particles that may either occur in the bulk as in the purely non-convective regime or near the boundary as in the convective regime. Such models have been also explored to explain size separation in flows in gravity driven inclined chute flows and silo flows~\cite{savage88}. 

\subsubsection{Void-filling models}\label{void_mod}

As particles can move or fluctuate about their positions upon vibration, one can note that the complementary void spaces between the particles also changes and fluctuates as the system is oscillated. For example, at the end of the upward stroke, spaces can open up as the grains gain kinetic energy and fly off the bottom boundary as it slows down and moves in the downward direction. Now as the particles collide and fall in gravity, the probability of the smaller particles filling a void is greater than a large particle. Thus, the idea goes, the large particle will end at a slightly higher position on an average at each step and therefore will rise to the top of the bed. This idea can be traced back to almost every early discussion on size separation in the presence of gravity. 

This model will imply a size ratio dependent rise-time for the large particle because it will enter into any estimate of the relative probability of particles filling the voids. One also notes that there is no critical size required for the rise of the large particles. The driving parameters can be anticipated to be important to the creation of the voids and thus also affect the rise-time. 

Before going on to making quantitative estimates, let us check if the idea by itself works qualitatively. After all voids are being created uniformly inside the bed at least as assumed in the simplest discussions. Rosato, {\em et al.}~\cite{rosato87} used Monte Carlo simulations to study this model of size separation. The shaking was simulated by repeatedly lifting the assembly of particles through the same height, and then allowing them to fall using the MC algorithm. In this step, a particle were allowed to either move down or horizontally through a random displacement provided it did not overlap with the position of other particles. They were able to show a large particle would rise because if a void opened up below it, the space was more likely to be filled by a small particle. Thus the large particle would be prevented from falling relative to the small particles and would gradually rise. A crucial ingredient is the anisotropic way in which particle positions are changed in the up and down direction because the kinetic energy is considered to be much smaller than the potential energy. Note that the particles are also considered to have essentially infinite densities. The presence of a particle, however small, will prevent a large particle from moving into that place. Furthermore small particles are not allowed to be squeezed out. Therefore mass dependence cannot be found in this realization. 

\begin{figure}
\begin{center}
\includegraphics[width=.5\textwidth]{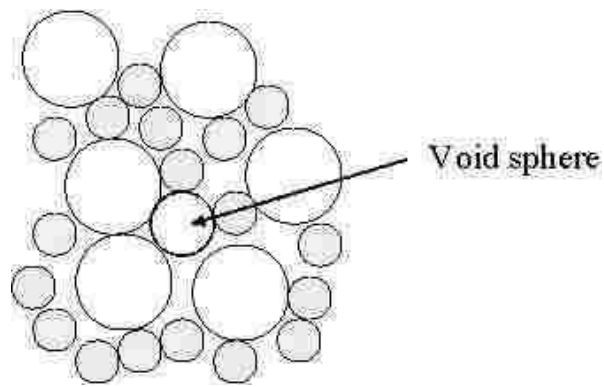}
\end{center}
\caption{A schematic of a layer composed of binary distribution of particles. A void sphere of diameter $d_v$ is also denoted. Adapted from Ref.~\cite{savage88}.}
\label{void_fill}
\end{figure}

A slightly more general formulation of this mechanism can be made without any reference to the shake cycle by following the work of Savage and Lun~\cite{savage88} in describing size-separation in chute flows. They followed what they called an maximum-entropy approach and first considered a binary mixture of particles in one layer as in Fig.~\ref{void_fill}. By considering the number of ways in which the particles and the void of diameter $d_v$ can be rearranged, they derived a percolation velocity of the two kinds of particles as a function of $r_d$ in the limit of a small number of large particles. In this formalism, each shake would be thought of as giving rise to a new arrangement of the kind shown in Fig.~\ref{void_fill}. 

If shaking amplitude is small or the particles are very densely packed as in a closed packed case, then the assumptions of sampling all possible configurations is likely to fail, and may be the cause of the lack of intruder motion seen in some experiments with small size differences (see Fig.~\ref{rise_time}). Nonetheless, this work represents one of the few attempts to derive quantitative percolation rates from system parameters and needs to be pursued further.

While the void-filling model neglects convective motion, the intruder location in the convective regime can be qualitatively understood by using this geometric model. When the large particle reaches the boundary where a thin flowing region exists, small particles move into the voids created by departing  particles. Thus while large particles may be carried to a certain location (be it at the top or the bottom) due to convection~\cite{knight93}, the reason they get stuck is due to the higher probability of the voids-filling by smaller particles. When the size difference $r_d$ is large, the probability for large particles to fall through the voids created in the thin boundary layer appears to reduce to zero. A recent complementary discussion can be also found in Ref.~\cite{rosato02}.

\subsubsection{Arching models}\label{crit_size}

Jullien, {\em et al.}.~\cite{jullien92} introduced an alternate geometrical model based on the differences between deposition of large and small particles during a ``shake" cycle. In their three dimensional simulations, the particles are lifted uniformly then dropped sequentially {\em starting with the lowest particle} along a random trajectory onto the substrate. The particle would move along the path of steepest descent after hitting a particle in the heap until it came to rest when it was supported either by the substrate or when the center of the particle was within the triangle formed by the vertical projections of the center of the three contacting particles. Now a large particle could shelter a large void below it from small particles which rain down later. In the next sequential update, the small particles which are below the center of the large particle, can move into the void thus preventing the large particle from falling. Jullien, {\em et al.}~\cite{jullien92} found in their simulations that large particles rise provided $r_d > 2.8$. 

Some experimental reports of critical $r_d$ in quasi-two dimensional experiments conduced at low vibrational amplitudes have been reported~\cite{duran93}. For example, a critical size for size separation appears to be implied if one extrapolates the trends seen for the rise-velocity shown in Fig.~\ref{rise_time}. Closed packing of the small particles was observed in both these experiments. A critical $r_d$ similar to that noted by Jullien, {\em et al.}~\cite{jullien92} was calculated by Duran, {\em et al.}~\cite{duran93} by analyzing the stability of a large particle in a tetrahedral (3D) or triangular (2D) packing. It is interesting to ask if such a threshold exists in the absence of close packing order as may be the situation away from the boundaries or with binary mixtures. 

The results for a critical size ratio is puzzling in many ways. Size separation well below the $r_d$ predicted by these models have been seen in many other experiments and simulations, although the conditions under which it is observed are not always clear. (This critical size cutoff is certainly not seen in binary mixtures as we will discuss in the next section.) In the limit of a small number of large particles, it is unclear if the specific sequential addition of the particles as implemented in Ref.~\cite{jullien92}, or the nature of the packing assumed in Ref.~\cite{duran93} is important to the observation of the threshold. An interested reader is referred to Refs.~\cite{barker93,jullien93,cooke96} for further discussion. 

\subsubsection{Other geometric models}

The link between geometric frustration in rearrangement and segregation was explored using a phenomenological Tetris-like model by Caglioti, {\em et al.}~\cite{caglioti98}. These ideas are interesting in the context of recent models of granular media based on configuration entropy models~\cite{edwards93}.\\

Now it is important to distinguish between the geometric mechanism for size separation and the size separation discussed in the vibro-fluidized limit. There the large particle moves to a position where the vertical pressure gradient matches its weight. Thus the mechanism is mainly a buoyancy driven separation where the location of the particle is given by the gradients present in the system. In the purely geometric limit considered here, particles are stochastically but uniformly driven, and the effects of the gradients are neglected. In principle, temperature gradients if not density gradients exist even for dense packing and may need to be evaluated on a case by case basis. 

\subsection{Effect of density}\label{intruder_density}

It has been repeatedly asserted that the size difference between particles and thus geometry is the most important factor in determining the intruder motion. However this does not mean that other properties such as density differences and other material properties are not important. Next, we will examine the effect of density on the motion of the intruder. 

In the vibro-fluidized regime first considered in this section, density can have a significant effect. For example, the simulations of Shishodia and Wassgren~\cite{shishodia01} show an almost linear decrease in the height of the intruder as $r_\rho$ is varied from 0.6 to 0.75. Considering the manner in which pressure and its gradients vary inside a vibro-fluidized bed, one may even anticipate a density at which the pressure is not strong enough to counteract gravity thus causing the intruder to sink to the bottom. Indeed, MD-simulations show that the intruder sinks smoothly to the bottom as $r_\rho$ is increased beyond one~\cite{ohtsuki95}. Few experiments have explored this presumably simple regime, but we will return to this point in the next section when we discuss binary mixtures. 

\begin{figure}
\begin{center}
\resizebox{7 cm}{5 cm}{\includegraphics{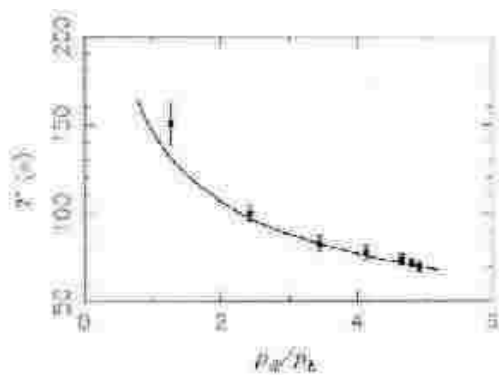}}
\end{center}
\caption{The rise time measured for an intruder by Liffman, {\em et al.}~\cite{liffman01} in a quasi-two dimensional experiment as a function of density ratio while keeping the size ratio fixed. Counter intuitively, the rise-time decreases with an increase in $r_\rho$. A curve corresponding to a theoretical model which predicts a $1/\sqrt{r_\rho}$ dependence is also shown.} 
\label{liff_den}
\end{figure}

Now let us consider a dense system and the role of particle density. Liffman, {\em et al.}~\cite{liffman01} undertook a systematic experimental study of an intruder in a quasi-two dimensional system to understand the role of intruder density in the non-convective regime. They used hollowed steel disks to vary the weight and thus relative density compared with the smaller glass particles. The rise time as a function of initial depth, disk size and density, and vibration frequency was reported. (Interestingly, it was noted that changes in the humidity in the room caused an order of magnitude changes in the measured rise times. Therefore the experiments were performed in an air conditioned room to reduce effects of humidity. This observation, however should warn us to compare across experiments where such details have not been noted.) Their measured effect of density on the rise time is plotted in Fig.~\ref{liff_den}. Somewhat counter-intuitively, the rise time increases with a decrease in density ratio. We will analyze the density dependence in more detail next, but note here that the mass dependence clearly demonstrates the importance of kinetics even in the deep bed limit. 

Liffman, {\em et al.}~\cite{liffman01} then developed a model for the size and density dependent rise. They assumed that each time the container is oscillated, the intruder is ``thrown" through the granular bed, and it is subsequently brought to rest due to a granular friction force. They noted that during the upward motion of an intruder, a wedge of material above it is displaced in two dimensions (a cone in three dimensions) due to force chains which run through the bed and terminate at the free surface. On the other hand, during downward motion, the force chains would terminate at the bottom plate and therefore give an effectively infinite mass. (Now it is unclear if this is true because to some extent the bed gets lofted off the base plate.) Next by considering the geometry and the density of the two species, they derived the ascent velocity as a function of depth $z$ from the surface which can be integrated to give the rise time $T_r$: 
\begin{equation}
T_r = \frac{F}{\Gamma d_L} \sqrt{\frac{1}{r_\rho d_S}},
\end{equation}
where, $F$ is a function of initial depth, friction, and other material parameters. Their model predicts that the total time for the disk to reach the surface should scale with the disk size, and should be inversely proportional to the square root of the disk density. This is compared to experimental results in Fig.~\ref{liff_den}. It is interesting to see that their model captures the upward trend of the rise with decreasing $r_\rho$. For $r_\rho \sim 1$, the theory under predicts the observed rise time. We will return to this discrepancy when we discuss the effect of interstitial air in Section~\ref{air_eff}. They further derived corrections to the above form near the surface and found good agreement with the measured position of the intruder as it ascended in time. The observed frequency dependence of the rise velocity was also observed be consistent with their model. 

At this point it is also interesting to note that all experimental observations and theoretical models {\em in the dense limit} appear to suggest that the intruder will always rise to the top. It is as yet unclear if this is simply because a large enough parameter space has not been explored or if it points to something fundamental.  Nonetheless it is clear that the monotonic rise with density is not true in the vibro-fluidized limit, where dense particles do sink. We will discuss this issue further in Section~\ref{bi_den} on density effects in binary mixtures, where there are experimental results in this limit as well.  

Now let us consider the opposite case when the intruder is less dense. In experiments performed with an intruder in a deep bed, Shinbrot and Muzzio~\cite{shinbrot98} discovered that when the density of the intruder was lighter than the surrounding particles, it sank to the bottom of the container displaying what they called reverse buoyancy. When $r_\rho \sim 1$, the intruder performed periodic convection motion. Whereas for $r_\rho > 1$, the intruder ascends and the rise times  were qualitatively similar to that shown in Fig.~\ref{liff_den}. They gave an explanation for the rising of the intruder in terms of inertial effects, but also commented that the impact of the entrained air on the observed reverse buoyancy needs to be explored~\cite{shinbrot98}.

\subsection{Effect of interstitial fluid}\label{air_eff}

The effects of interstitial air have been investigated only recently and shown to dramatically affect the motion of the intruder~\cite{mobius01,yan03}. Mobius, {\em et al.}~\cite{mobius01} appear to indicate that interstitial air plays a prominent role in the motion of a light intruder. A non-monotonic dependence of the rise time was found as a function of density ratio of the intruder to the surrounding particles. Consistent with previous observations, the rise time first increased as the $r_\rho$ was decreased. As $r_\rho$ was decreased further below one, it was observed that rise time then decreased. By evacuating the interstitial air from the container, they observed that the overall variation of the rise time changed significantly at low $r_\rho$. Both the amplitude and the location of the peak in the rise time was observed to decrease considerably when the air in the container is evacuated. Moreover, Mobius, {\em et al.}~\cite{mobius01} did not find a regime where the large particle sank in their experiments. 

\begin{figure}
\begin{center}
(a)\resizebox{7 cm}{5 cm}{\includegraphics{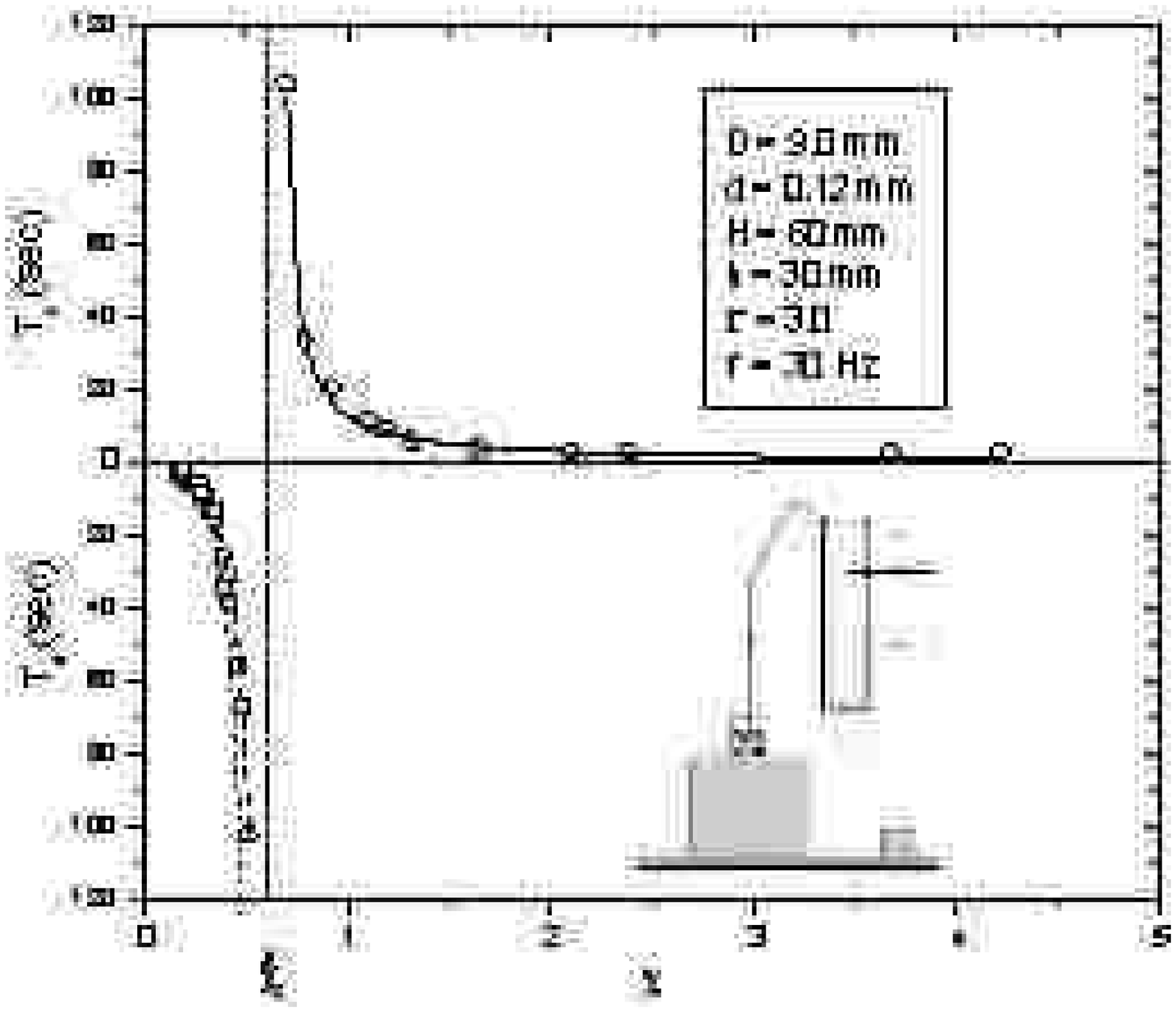}}
(b)\resizebox{7 cm}{5 cm}{\includegraphics{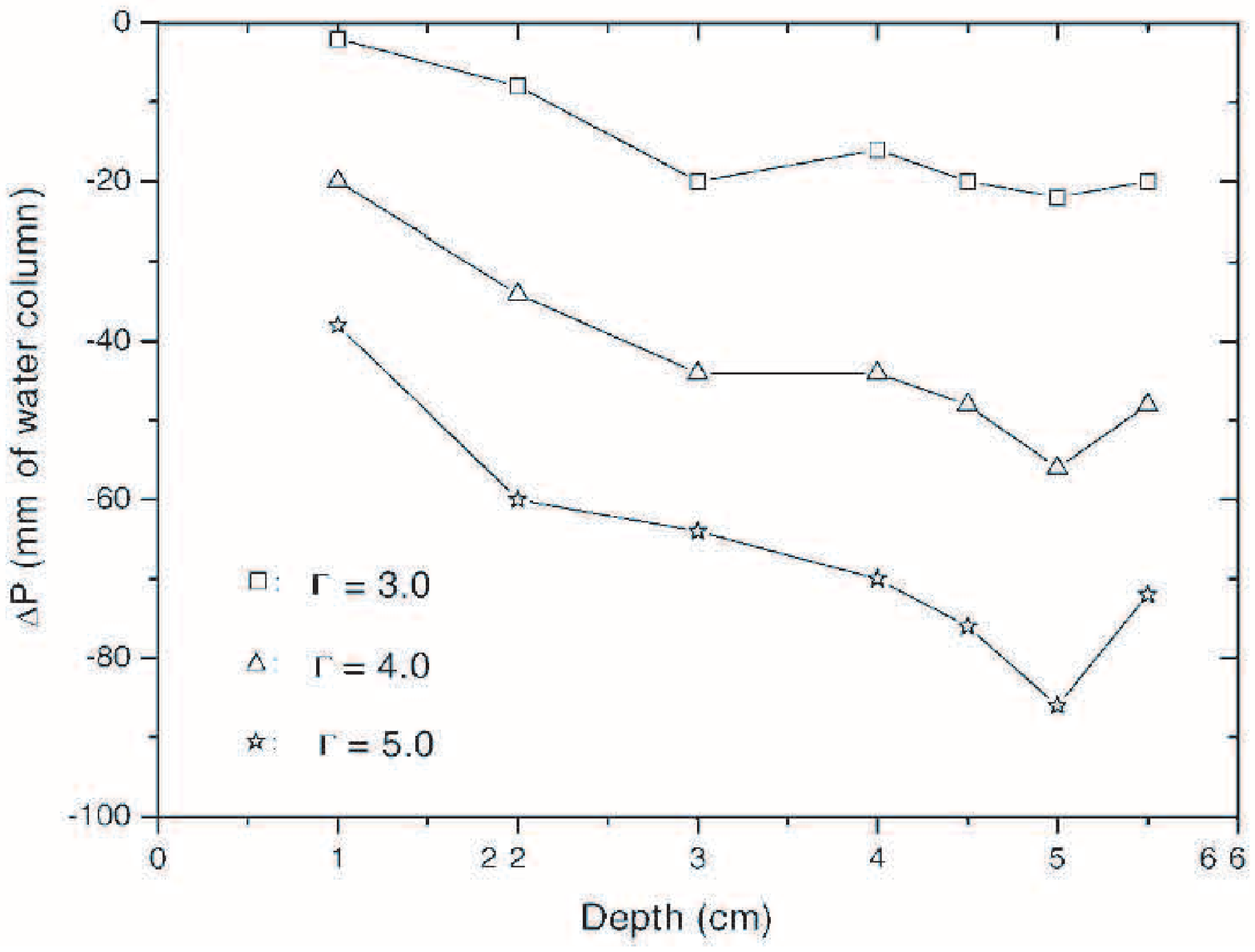}}
\end{center}
\caption{(a) The rise or sink time of the intruder as a function of the density ratio of the intruder and the small particles. It is observed to increase rapidly at a critical $r_\rho$ independent of particle size. (b) Air pressure variation measured inside the axis of a bed by Yan, {\em et al.}~\cite{yan03}. The pressure variation inside the container can significantly affect the motion of an intruder when $r_\rho < 1$. From Ref.~\cite{yan03}.} 
\label{air_eff_fig}
\end{figure}

In detailed experiments performed recently by Yan, {\em et al.}~\cite{yan03}, the effect of interstitial air was further investigated. In particular for a large particle in a bed which is vibrated vertically, they found that the intruder may rise up or sink depending on $r_\rho$ and $r_d$ in the presence of interstitial air [see Fig.~\ref{air_eff_fig}(a)]. However, the intruder was always found to rise when the interstitial air in the container was evacuated. The pressure variation inside the container were measured and was found to decreases as a function of depth inside the container [see Fig.~\ref{air_eff_fig}(b)]. The pressure variation depended on the size of the small particles and was most prominient for small bed particles. Therefore it was concluded that the gradient of the air pressure across a large particle causes it to sink. However, further experiments and analysis are necessary to understand the relative importance of air pressure difference which apparently cause a large particle to sink, and the mechanisms that cause the large particle to sink.  

\section{Size separation in binary mixtures}

Now we will discuss size separation when the volume fraction of the large and small particles are comparable, and interactions between the large particles introduce crucial differences from the intruder model system. This is also the regime where almost all size separation is encountered in a material processing application. We will start this section with a discussion of binary mixtures composed of the same material but different sizes, and compare and contrast the results with the intruder model system. Then, we will consider additional physical factors that impact size separation as in the previous section.

\subsection{Mixing-demixing transitions}

\begin{figure}
\begin{center}
(a)\includegraphics[width=.45\textwidth]{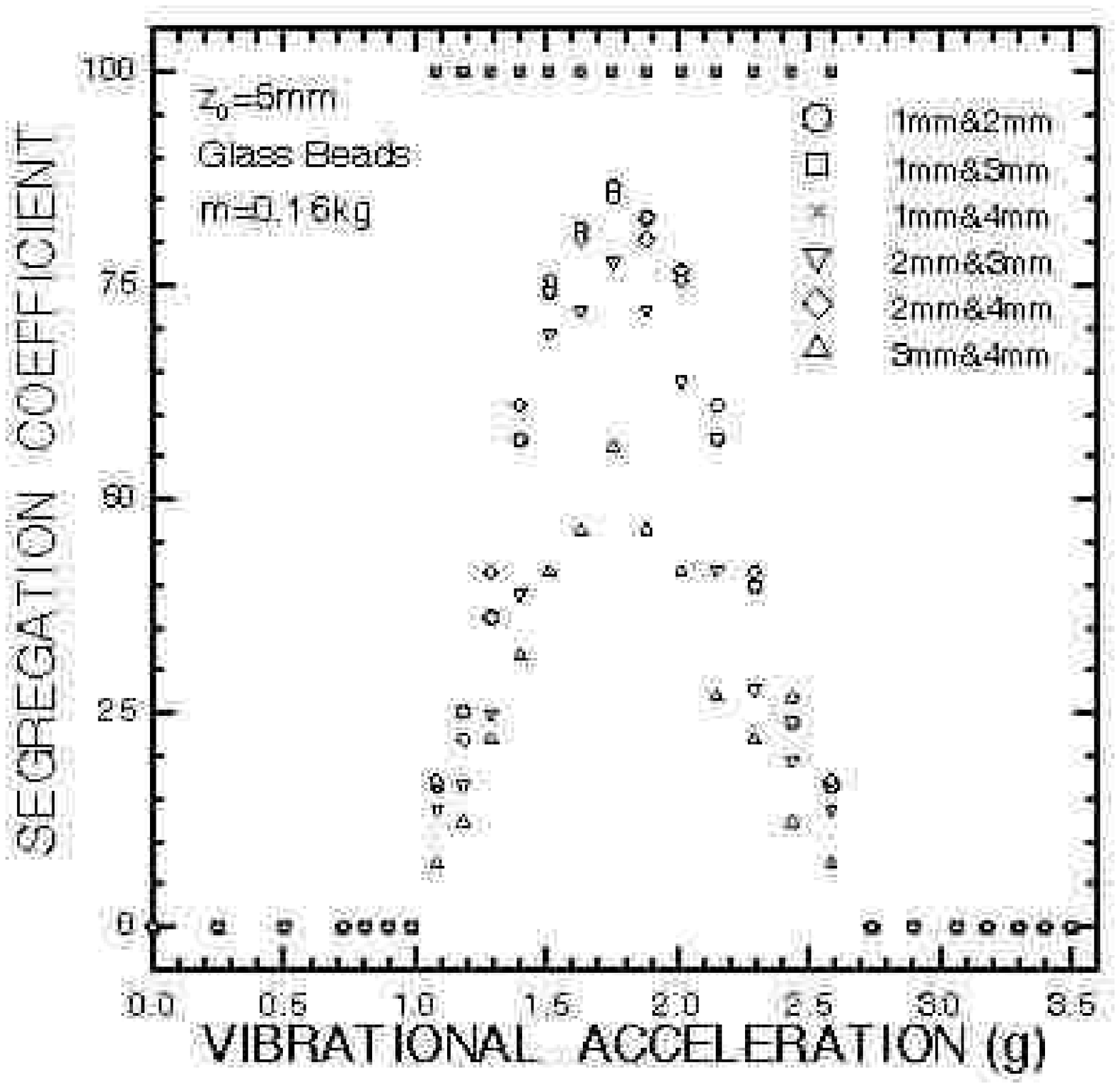}
(b)\includegraphics[width=.41\textwidth]{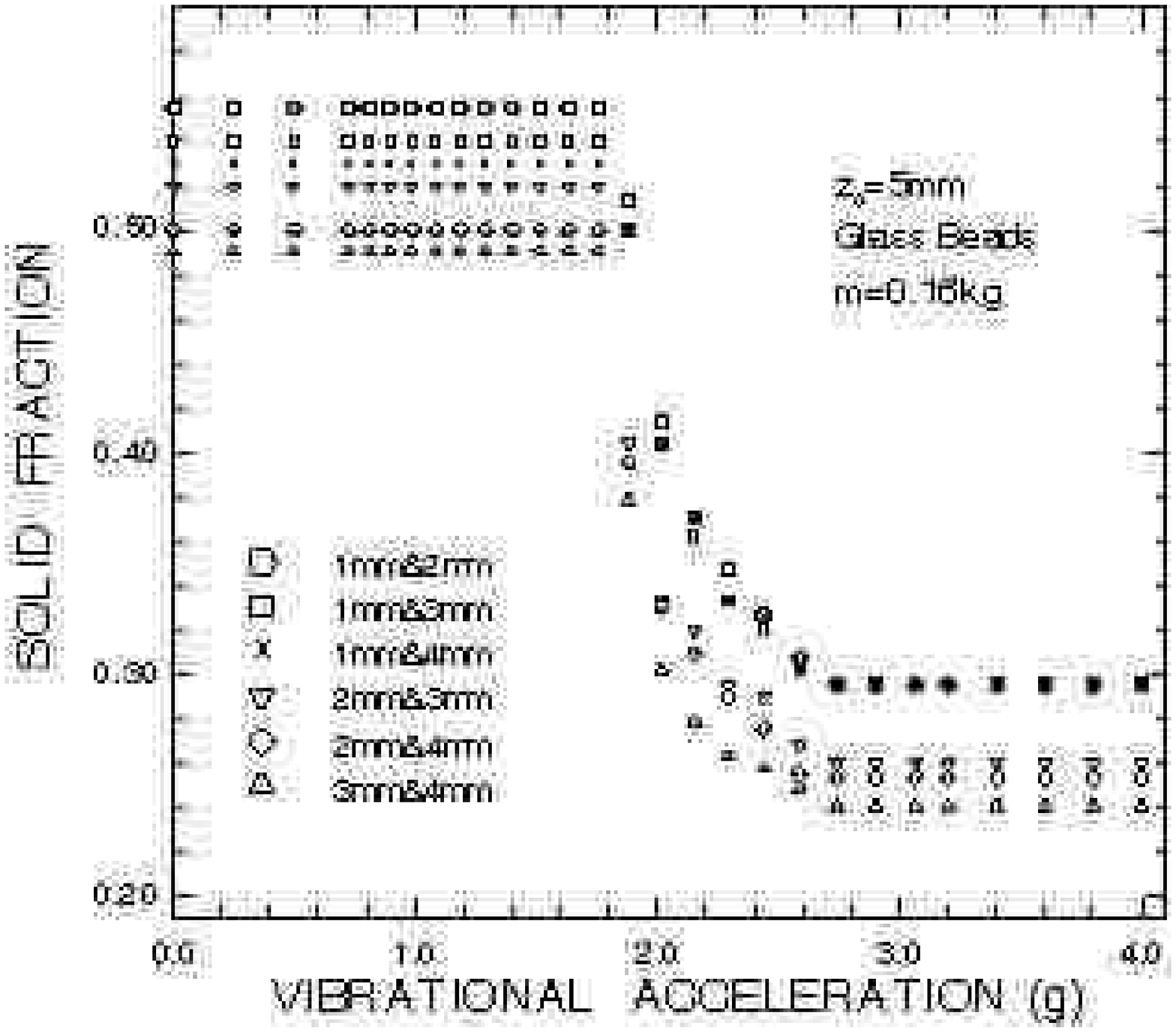}
\end{center}
\caption{(a) Segregation in binary mixtures as a function of vibrational amplitude measured by Hsiau and Yu~\cite{hsiau97}. (b) The corresponding mean solid fraction of the system. The segregation is observed to increase till $\Gamma = 1.75$ but then decrease completely to zero when the system is highly fluidized.} 
\label{hsiau_seg}
\end{figure}

Let us first consider a container filled with a nearly equal volume fraction of large and small particles of the same kind, which is being vibrated in a container where the side wall and the interstitial fluid can be ignored. Such a system has been systematically investigated experimentally by Hsiau and Yu~\cite{hsiau97,hsiau00} using glass and steel beads in a plexiglass cylinder ($r_d = 1.25, 1.33, 2, 3, 4$, $D = 20$ cm). Because a large number of both species of particles are present, one has to discuss spatial distributions rather than discussing the position of the large particle as we did in the previous section. Hsiau and Yu visualized the mixture as a function of vibrational acceleration and measured the ratio of the difference of the number of large particles in the top half and the bottom half of the container to their total number. This parameter was used as a quantitative measure of segregation and is shown in Fig.~\ref{hsiau_seg}. It was found that segregation first increased with amplitude as the system was fluidized above $\Gamma =1$. However, as the solid-fraction of the bed decreases, segregation also decreases rapidly and the two species were more or less completely mixed. It can be also noted from Fig.~\ref{hsiau_seg} that segregation occurs even for $r_d$ as small as 1.25 (much lower than the $r_c = 2.8$ discussed in Section~\ref{crit_size}).

Qualitatively similar experimental results were found earlier by Rippie and collaborators in a somewhat smaller system~\cite{olsen64,rippie64}. In order to quantify the degree of segregation, they found the standard deviation of the system by measuring the weight ratio of the components in six horizontal slices (rather than two halves as in the experiments of Hsiau and Yu). Starting with a random sample, they found that the standard deviation changes exponentially with time and thus concluded that the segregation process is first order. They also found that the segregation rate (calculated from the exponential behavior) was almost linearly proportional to $r_d^3$ and depended on the overall volume of the two components. They also claimed that the side walls did not significantly affect the segregation rates provided the cylinder diameter was large enough and convection was absent. 

The qualitative trends observed in these experiments can be explained in terms of the void-filling mechanism discussed in Section~\ref{void_mod}. As the container, filled with an initial mixed distribution of particles, is vibrated at small amplitudes, particles vibrate and acquire energy to fall into voids that are created between the particles. As discussed in Section~\ref{void_mod}, the probability of smaller particles falling into a void is greater compared to the larger particles. However, as the shaker is vibrated more strongly, the particles acquire a lot more energy and the system gets strongly vibro-fluidized, the voids increase in size to such an extent that the relative probability between large and small particles to fall into a void decreases. Thus segregation decreases. There have been recent developments showing that the center mass of the system  increases as the square of the vibrational velocity of the bottom plate (and thus overall compaction decreases)~\cite{luding94,kumaran98}. However, these results are for dilute system and will have to be extended to dense systems before a more quantitative estimate of the progress of size separation with $r_d$ and $\Gamma$ can be carried out. 

\subsection{Effect of convection on segregation and mixing}\label{binary_conv}

As discussed in Sections~\ref{convection}~and~\ref{seg_con}, friction with side walls and interstitial air can drive convection in a vibrated granular bed and impact the motion of a large particle. Therefore convection can be anticipated to affect segregation in binary mixtures. In fact, an argument can be made based on intuition from normal fluids, that convection can enhance mixing and consequently counteract the effects of mechanisms which cause segregation. 

In vibrated granular systems, there is significant support for this expectation. Ratkai~\cite{ratkai76} proposed a novel experimental design in which only the central region of the bottom boundary is vibrated. As demonstrated by Ratkai, this type of driving promotes an upward flow in the center and a downward flow at the boundary, thoroughly mixing materials of two kinds which were initially placed in a layered pattern inside the container~\cite{ratkai76}.

\begin{figure}
\begin{center}
\includegraphics[width=.7\textwidth]{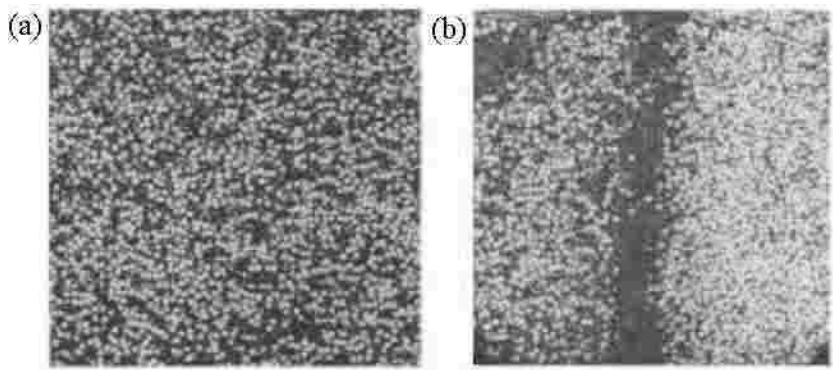}
\end{center}
\caption{(a) An image of a vertical slice of the bed obtained after a binary mixture is vibrated. Binders are introduced to hold the particles in place after the mixture is vibrated to reveal the structures. Convection was observed to destroy up-down size separation by Brone and Muzzio~\cite{brone97} in experiments with binary mixtures at high vibration frequencies. (b) A radial size separation was observed when heaping occurred at low driving frequencies.} 
\label{brone_seg}
\end{figure}

Using a normal rigid vibrated container, Brone and Muzzio~\cite{brone97} explored the interplay between convection, size separation and mixing. Binders were used to freeze the particles in place and then slice the solidified structures to reveal the internal structures. At high $\Gamma$, they found that while a single large particle in a bed of small particles rose to the top as described by Knight, {\em et al.}~\cite{knight93}, a homogeneous spatial distribution was observed with a 50-50 binary mixture [see Fig.~\ref{brone_seg}(a)]. The bed surface was flat in this regime. 

However, at lower $\Gamma$ {\em radial size separation} was observed inside the container with large particles at the sides and small particles in the center. Heaping was also observed in this regime with the center being higher than the sides [see Fig.~\ref{brone_seg}(b)]. They explained the radial distribution by noting differences between the rapid motion of the large particle and the slow motion of small particles at the free surface which is inclined with respect to the horizontal due to the formation of the heap. Thus it appears that these radial patterns have their origin as the segregation on inclined chute flows due to the void-filling mechanisms~\cite{savage88}.  

In the experiments of Brone and Muzzio, it is not apparent if interstitial air besides boundary friction is important in promoting convection. Furthermore it is possible that the relative importance of the factors that promote convection might have also changed as the frequency was varied. Nonetheless it is clear that the size separation phenomena in binary mixtures is qualitatively different than observed in the intruder system. 

To understand if mixing will indeed occur in a situation as in the experiments of Knight, {\em et al.}~\cite{knight93}, let us consider a situation where a large particle travels to the top and stays there because it is too large to enter the thin boundary layer with downward moving particles. If one keeps adding a large particle to the initial mixture, one will reach a point eventually where a sizable number of large particles will accumulate at the top and near the boundaries. Unless the frictional property of the large particles is considerably different than the smaller ones, it has to be assumed that the large particles themselves will feel a phase dependent downward force~\cite{grossman97} due to interaction with side walls. Thus the large particles could also eventually move down the side walls. It is unclear as to what would exactly happen next or at what volume fractions the effect would kick in, but it is clear that for a comparable volume fraction of large and small particles, considerable mixing is likely. It is possible that novel segregation mechanism in the thin downward flowing shear zone may occur, but a significant number of investigations have not yet been accomplished for a detailed picture to emerge. 

Few simulations have been performed in the dense limit where interaction with side walls is significant because of the  frequent interactions among particles and boundaries. In one of the few simulations performed in this regime, Elperin and Golshtein~\cite{elperin97} observed subtle effects of convection and size separation. In their MD simulations, it was found that large particles can migrate to the center of the convection rolls which form in the container. However, the exact mechanism for this novel spatial distribution is not clear, nor has a similar observation been reported in experiments.  

\subsection{Density effects in binary mixtures}\label{bi_den}

Up until now we have used only geometric arguments to explain spatial distributions in binary mixtures. Now let us examine the effect of density ratio $r_\rho$, which would imply that kinetic effects are also important. In the pioneering experiments of Rippie, {\em et al.}~\cite{rippie64}, segregation rates were found to be greater when the smaller particles were denser. These trends appears to contradict the observations discussed in Section~\ref{intruder_density}, where we saw that the rise time of an intruder decreases with an increase in $r_\rho$. This once again underscores the point that the intruder model system and binary mixtures can show considerably different behavior. However, there is not much data available to carry out further discussion on segregation rate dependence on $r_\rho$. 

Therefore let us continue on and examine how particle density affects the spatial distribution in binary mixtures. Using MD simulations, Hong, Quinn, and Luding~\cite{hong01} showed that it is possible for large particles to sink to the bottom of a homogeneously excited binary mixture which differed in size and density, provided the large particles were dense enough. They called this the reverse Brazil nut effect. Their simulations were conducted in two and three dimensions with particles with a coefficient of restitution of $0.9999$ for the particles and $0.98$ for the walls. Furthermore, rather than using an oscillating boundary to supply energy, particles were uniformly given energy in space using a random driving which mimics a thermal bath. Following our discussion in Section~\ref{back}, it may be noted that this situation is considerably more idealized than in an experiment. Hong, {\em et al.}~\cite{hong01} also derived a criterion for when large particles would rise or sink depending of $r_\rho$ which we will discuss next. 

\subsubsection{The condensation model}

Hong, {\em et al.}~\cite{hong01} proposed that a competition between condensation and percolation in mixtures determines the spatial distribution of particles which differ by size and density. They first argued that a condensation transition exists for particles in gravity below a critical temperature~\cite{hong99,quinn00}, which depends on the mass and size of the particle. At higher temperatures, particles would be fluidized and display Boltzmann statistics because they were in contact with a uniform thermal reservoir. It was then argued that if this system is in contact with a bath with a temperature which is in between the critical temperatures for the two species, then it would be possible for large particles with high enough denisty to condense at the bottom. They determined the criterion as:
\begin{equation}
c_h = \frac{r_\rho}{r_d^{D_s-1}},
\label{hong_cri} 
\end{equation}
where $D_s$ is the spatial dimension and the large particle would rise if $c_h > 1$, or sink otherwise. The criterion appears simple enough but it is important to note some key assumptions that have been made in deriving this criterion.  First, particles are assumed to be {\em elastic}. Second, inter-species interactions are completely neglected. Third, the temperature in the system was assumed to be homogeneous. And finally, the temperature of the two species is assumed to be the same. 

A number of issues~\cite{walliser02,canul02,quinn02} were raised with the findings of Hong, {\em et al.}~\cite{hong01}. Canul-Chay, {\em et al.}~\cite{canul02} tried to find experimental evidence for the criterion proposed by Hong, {\em et al.} While they found states in which the large particles were on top at low densities, only mixed states were observed when the density of the large particles was greater than the small particles. 

\begin{figure}
\begin{center}
\includegraphics[width=.5\textwidth]{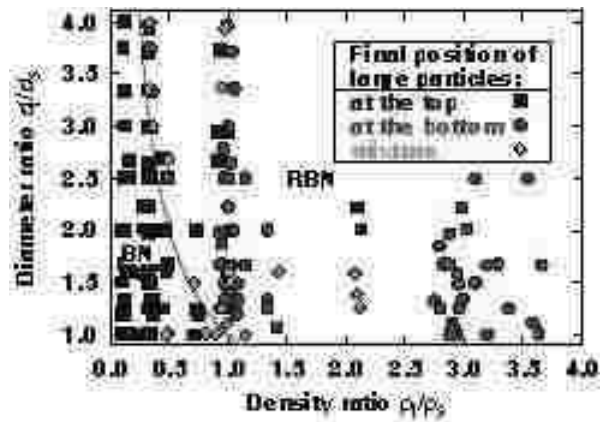}
\end{center}
\caption{The final position of the large particles as function of size ratio $r_d$ and density ratio $r_\rho$. BN: Brazil Nut; RBN: Reverse Brazil Nut. From Ref.~\cite{breu03}.}
\label{den_phase}
\end{figure}

More recently, Breu, {\em et al.}~\cite{breu03} have experimentally investigated conditions under which the large particles sink to the bottom (see Fig.~\ref{breu_fig1}). They found spatial distributions depending on the density ratio as well as on external conditions. The phase diagrams as a function of diameter ratio and density ratio were reported (see Fig.~\ref{den_phase}). Furthermore, the effect of normalized acceleration and frequency was investigated. They claimed that the experiments confirmed the theory of Hong, {\em et al.}~\cite{hong01} provided a number of conditions were chosen carefully. These factors included using fairly rigid particles, and choosing an intermediate vibration amplitude. In addition, granular temperature was not measured directly and therefore no firm conclusions can be drawn as to the validity of the condensation model from the published data. 

It is interesting to ask if the reason why Breu, {\em et al.}~\cite{breu03} could observe large particles at the bottom was because they were in the vibro-fluidized limit, whereas the experiments of Canul-Chay, {\em et al.}~\cite{canul02} were not. On this point at least we are back to almost where Brown~\cite{brown39} left off in his first discussions on the interplay between size and density. While it appears that heavy particles sink in the vibro-fluidized limit when particles undergo binary collisions, it is not entirely clear if the same is true in a regime where enduring contacts occur as in deep beds. The answers are not available currently and these recent investigations if sustained may be able to provide them. 

\subsubsection{Kinetic model of binary mixtures} 

The kinetic approach described in Section~\ref{kinetic} was extended to binary mixtures~\cite{jenkins87} and was later applied to the issue of segregation in vibrated systems by Jenkins and Yoon~\cite{jenkins02}. As discussed in Section~\ref{kinetic}, the particles are assumed to be smooth and therefore frictionless. This means that rotational degrees of freedom and energy loss in the tangential direction during a collision are ignored. Assuming that the temperature is uniform, they derive a criterion for which species will rise initially as a function of their density and size ratio. The Jenkins-Yoon criterion for a small number of large spheres reduces to: 
\begin{equation}
c_{3D} = \frac{24 r_\rho}{(r_d +1)(r_d + 2)(r_d + 3)}
\label{c3d}
\end{equation} 
where, the large particles would rise for $c_{3D} > 1$, and sink otherwise. A similar criterion was also obtained for disks~\cite{jenkins02}. We note that the criterion is for initial segregation. It is unclear what the steady state criterion is likely to be. 

The physical reason for the segregation apparently is due to the inertia of the particles. The prediction of Jenkins and Yoon is somewhat similar to the prediction of Hong, {\em et al.}~\cite{hong01}, however the mechanisms are apparently very different. They claimed that the obtained segregation criterion Eq.~\ref{c3d} for spheres and disks were supported by numerical simulations discussed in Ref.~\cite{hong01} because assumptions in the simulations were the same. 

\subsubsection{Hydrodynamic model assuming binary temperatures}

As noted in Section~\ref{binarytemp}, the temperature of two species in a vibrated system may not be the same. This was considered by Trujilo and Herrmann~\cite{trujillo02} while using the kinetic approach. In their model for a dilute system of large particles (intruder model), the local density is modified because the presence of a large particle making the region locally hot or cold. This leads to change in the effective local density, which according to their calculation, leads to a buoyancy force. They also derive a criterion for when an intruder would rise depending on its size and density. However, they assume that the pressure and the temperature are constant in the absence of the intruder. Therefore, the importance of temperature gradients which exist in the vibro-fluidized limit remains unresolved when energy is injected into the system at the boundaries. 

\subsection{Friction and inelasticity effects}

It must be noted that at least in experiments such as those by Breu, {\em et al.}~\cite{breu03}, materials with different densities may not also have the same friction and inelasticity. Similar results were found when glass or steel beads of various size ratios were used~\cite{rippie64} i.e., surface friction and density appears to not have a significant effect on segregation rates if they are only slightly different. However, no segregation was observed when lead particles with $r_d$ as high as $4.8$ were used. They attributed this effect to the considerably lower coefficient of restitution of the lead particles. They proposed that under the same excitation conditions, lead particles were not receiving enough energy to move relative to one another and therefore no segregation was observed.

In experiments with mixtures of angular shaped and smooth particles, Rippie, {\em et al.}~\cite{rippie67} found that convection was set up in their vibration experiments which was absent when just smooth or spherical particles was used. The convection gave rise to novel time dependence of spatial distributions in their system. However, many more investigations exploring the effect of these important material parameters need to be carried out. 

\section{Impact of direction of vibration}\label{vib_direc}

We have essentially considered vertically vibrated systems in most of this review, and now will briefly consider the effect of driving direction on size separation. In section~\ref{back}, we discussed that the energy injection from the boundaries can cause anisotropy in the particle fluctuations, and interactions with the side walls during a driving cycle can cause convection which impacts size separation. Therefore the direction of vibration can be anticipated to have considerable impact. 

There have been a few investigations using other kinds of vibrations including horizontal oscillations, shaking in horizontal and vertical circles and its variations. In industrial situations, all these various kinds of vibrations are used based on the application. The published literature on systematic studies of size separation is relatively small in these other kinds of vibrations and a detailed picture has not yet emerged on the influence of the driving. However, it is clear that size separation does occur even in horizontal shaking but the phenomena can be qualitatively different. 

It appears that experiments with horizontal shaking show a richer variety of strong convection patterns compared with purely vertical vibration. Even with a mono-layer of particles, convection is observed due to substrate interactions~\cite{painter00}. For a deeper bed, complex patterns have been observed in experiments and simulations depending on the depth of the bed~\cite{liffman97,metcalfe02}. 

When a bed of material is oscillated horizontally, the grains get pushed against the side in one part of the cycle, but a gap can open up in the other half. Based on detailed experiments with a bed of grains in a rectangular box, Metcalfe, {\em et al.}~\cite{metcalfe02} concluded that convective flow in the horizontal plane can occur because of side wall shearing. Furthermore, convection can occur in the vertical plane due to the avalanching of grains that occurs in the gaps that open up near the side wall as the grains get sloshed from side to side. Even more complex convection patterns are observed when a bed is shaken in the vertical as well as horizontal direction~\cite{tennakoon98}.

The complexity in the convection patterns gives rise to size separation that differs from vertically vibrated systems. In MD simulations performed in 2D (thus only vertical convection rolls are possible), Liffman, {\em et al.}~\cite{liffman97} showed that a few ``intruder" particles could go to the bottom of a horizontally shaken granular bed due to convection. Under similar simulation conditions, the particles would rise if shaken vertically. Mullin observed that the large particles in shallow layers with bi-disperse mixtures of particles cluster in stripe patterns orthogonal to the direction of driving~\cite{mullin00}.

\begin{figure}
\begin{center}
\includegraphics[width=.8\textwidth]{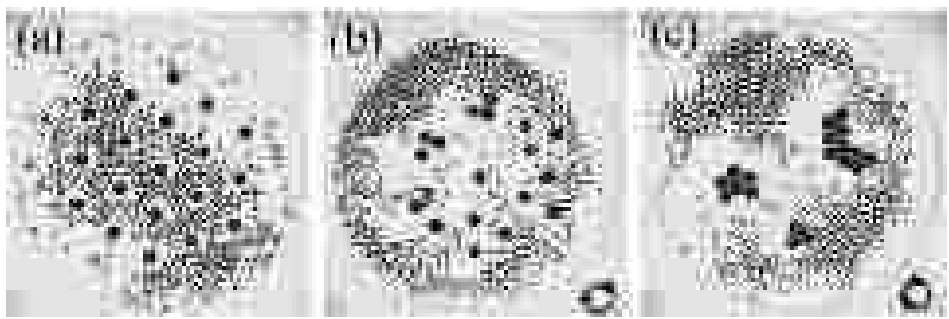}
\end{center}
\caption{Depletion induced segregation observed in a binary mixture of large disks and small spheres, which are swirled in horizontal circles. (a) t = 0 s, (b) t = 360 s, and (c) 4896 s. From Ref.~\cite{aumaitre01}.}
\label{depletion}
\end{figure}

Currently too little information exists to draw any broad conclusions about direction dependence and clearly much more work needs to done in order to understand the underlying flows. But a particular simple investigation by Aumaitre, {\em et al.}~\cite{aumaitre01} has important ramifications. An interesting propensity for the large particles to cluster was observed in a horizontally swirled mono-layer of large and small particles~\cite{aumaitre01}. A system of large disks and small spheres were swirled in a horizontal substrate, and over time large particles were observed to cluster together (see Fig.~\ref{depletion}). By tracking and hence measuring the velocity of the distribution of space they were able to estimate the pressure in the system as a function of position. They then showed that the pressure between two nearby large disks was lower on average than the mean. Thus the large disk would get pushed together and cluster. 

The explanation given for this remarkable effect is essentially the same as given to explain the phase separation in colloidal systems due to depletion forces. In equilibrium systems, excluded volume effects for small particles around hard-sphere large particles leads to an increase in entropy if two large particles are together~\cite{dinsmore95}. Thus entropically it turns out to be favorable for large particles to clump together leading to phase separation. It is interesting that these ideas appear to carry over to out of equilibrium systems, such as explored by Aumaitre, {\em et al.}~\cite{aumaitre01}. More of these kinds of studies are needed to compare and contrast the difference between size separation in and out of equilibrium. 

\section{Effect of cohesion on size separation}

Although we have assumed granular materials to be cohesionless up until now in the review, this is not always the case. Presence of humidity can affect size separation even in millimeter sized particles~\cite{williams76}. For example in the study of rise times of large particles by Liffman, {\em et al.}~\cite{liffman01} discussed in Section~\ref{intruder_density}, humidity in the room was observed to affect the rise time by an order of magnitude.  Furthermore, in a number of industrial process liquids are often added to reduce segregation effects. Addition cohesive interactions can exist in magnetized materials which can be found in the industrial processing of ores and toner for copying machines. We will therefore consider such effects in this section in the context of vibrated granular systems.   

\subsection{Impact of capillarity on segregation}

When a small amount of liquid is added to a granular mixture (or if humidity is present), liquid bridges develop between particles which introduce cohesive forces between particles due to the surface tension $\gamma$ of the liquid. If the size of the liquid bridge is large compared to the surface roughness of the particles, then the force of attraction between the particles is given by 
\begin{equation}
F_c = \pi \gamma d.
\label{capill}
\end{equation}
If however, the width of the bridge is small compared to roughness, then additional factors including the average asperity height, and the wetting angle have to be taken into account, as was discussed by Clark and Mason in the 1960s (see Refs.~\cite{clark67,mason68} for example). For typical glass or steel particles considered in this review with surface roughness on the order of a micron, Eq.~\ref{capill} appears to hold above 1\% by volume fraction of the liquid.  

The importance of cohesive forces to other forces can be judged by using the Bond number $Bo$ which is the ratio of the capillary forces to the weight of the particles~\cite{ottino00}: 
\begin{equation}
Bo = \frac{32 \gamma}{ 3 \rho d^2}.
\label{Bo}
\end{equation}
Therefore, for typical granular particle, $Bo$ can be of the order ten which gives one an idea of the size of a correlated clump in a wet granular mixture. Furthermore, $Bo$ depends inversely on the size of the particles which goes along with our intuition that smaller particles stick together better than larger particles. 

\begin{figure}
\begin{center}
\includegraphics[width=.9\textwidth]{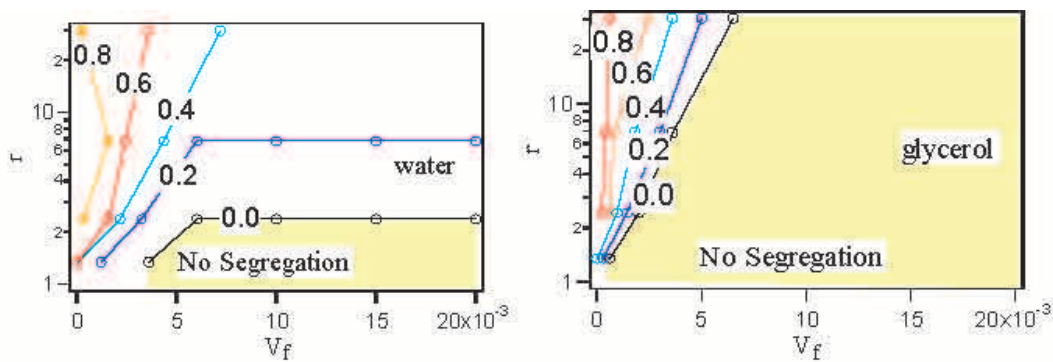}
\end{center}
\caption{(a) Size separation phase diagram as a function of size ratio $r$ and volume fraction of the added liquid (water) to particles $V_f$. The mixture is poured to form a triangular pile. Almost complete size separation is observed as a function of down hill position when the mixture is dry. A small amount is observed to completely kill size separation especially for small size ratios. (b) Higher viscosity of the fluid is observed to considerably diminish size separation. Adapted from Ref.~\cite{samadani00}.}
\label{sama_fig1}
\end{figure}

There are very few systematic studies of segregation in wet granular materials. Therefore we will need to use examples outside of vibrated systems to discuss some of the impact on the size separation properties. Recently, Samadani and Kudrolli~\cite{samadani00,samadani01} performed investigations of segregation in heaps as a function of size ratio of the particles and volume fraction of the added liquid (see Fig.~\ref{sama_fig1}). They demonstrated that a sharp decrease in segregation could be observed by adding as little as 1\% by volume fraction of the liquid, and that the decrease of segregation changed along with the cohesivity as measured by the angle of repose of the pile. 

In addition, viscosity of the liquid was also shown to play an important role in determining the properties of wet granular materials~\cite{samadani01}. The viscous force between particles which always acts opposite to the relative motion of the particles is given by: 
\begin{equation}
F_\nu = \frac{S_b \nu v_r}{\delta}
\end{equation}
where, $\nu$ is viscosity of the fluid, $v_r$ is relative velocity, $\delta$ is the separation, and $S_b$ is a factor which depends on the width of the bridge and the wetting angle between particles. These kind of forces, if present, suppress relative motion between particles and therefore counteract the void-filling mechanism and thus decrease segregation~\cite{bridgwater78,samadani00,samadani01}.

\begin{figure}
\begin{center}
\includegraphics[width=.45\textwidth]{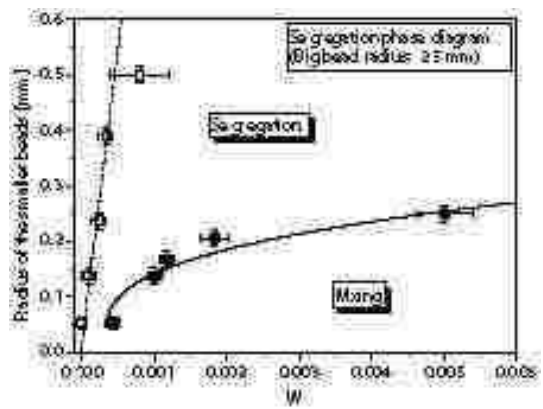}
\end{center}
\caption{The segregation phase diagram as a function of $W$ the ratio of the volume of the liquid and the volume of small beads. Size separation can be enhanced by adding a small amount of liquid to a binary granular mixture which is shaken in horizontal circles. The complex convection flows set up in this geometry should be taken into account before applying the results to the vertically vibrated case. Note that the phase space of greater size segregation is opposite to that shown in Fig.~\ref{sama_fig1}. From Ref.~\cite{gero03}.}
\label{wet_cohe}
\end{figure}

Now, as can be noted from Eq.~\ref{Bo}, there is an effectively stronger cohesive force between smaller particles than larger particles, which can lead to interesting size dependent interactions in the presence of liquid bridges. Samadani and Kudrolli noted that this may lead to layering in poured heaps. More recently, Li and McCarthy~\cite{li03} showed that by coating one set of particles with hydrophobic and one set with hydrophilic particles, {\em segregation could be induced by adding a liquid such as water}. 

Recently, Geromichalos, {\em et al.}~\cite{gero03} have also demonstrated segregation-mixing transition in wet granular mixtures shaken in horizontal circles. They found that for certain $r_d$, segregation can be enhanced by adding a small amount of liquid. They argued that a relative decrease in the coefficient of restitution leads to enhanced size separation. Given the complexity of the flows that arise during horizontal shaking, as discussed in Section~\ref{vib_direc}, this becomes a complex system to analyze. From the reported measurements it is unclear if mean flows are observed in their experiments and how they may change with volume fraction of the fluid added. 

\subsection{Magnetic and electrostatic interactions}

In addition to liquid induced interactions, electrostatic and magnetic forces can also be present which can induce cohesive interactions and jamming. For small particles it is common to observe adhesion to the side-wall due to screening interactions. There are no systematic studies available to gauge the importance of such effects on size separation. Therefore, we will conclude the discussion of cohesive interaction on size separation by discussing the clustering transition observed in a mixture of magnetized and non-magnetized beads in a vertically vibrated container~\cite{blair03b}. 

The idealized interaction between two magnetic or electric dipolar hard spheres separated
by distance $r$ is defined as
\begin{equation}
U_D = \frac{1}{r^3}  (\vec{\mu_i} \cdot \vec{\mu_j}) -
\frac{3}{r^5} (\vec{\mu_i} \cdot \vec{r}_{ij})(\vec{\mu_j} \cdot
\vec{r}_{ij}),
\label{eq:dhs}
\end{equation}
where, $\vec{\mu}$ is the dipole moment, and $\vec{r}$ is the  inter-particle vector connecting the centers of dipoles $i,j$. If the potential is averaged over all possible dipole arrangements, the effective potential is $r^{-6}$, similar to van der Waals interaction and is long ranged~\cite{degen_pinc}. 

\begin{figure}
\begin{center}
(a)\includegraphics[width=.5\textwidth]{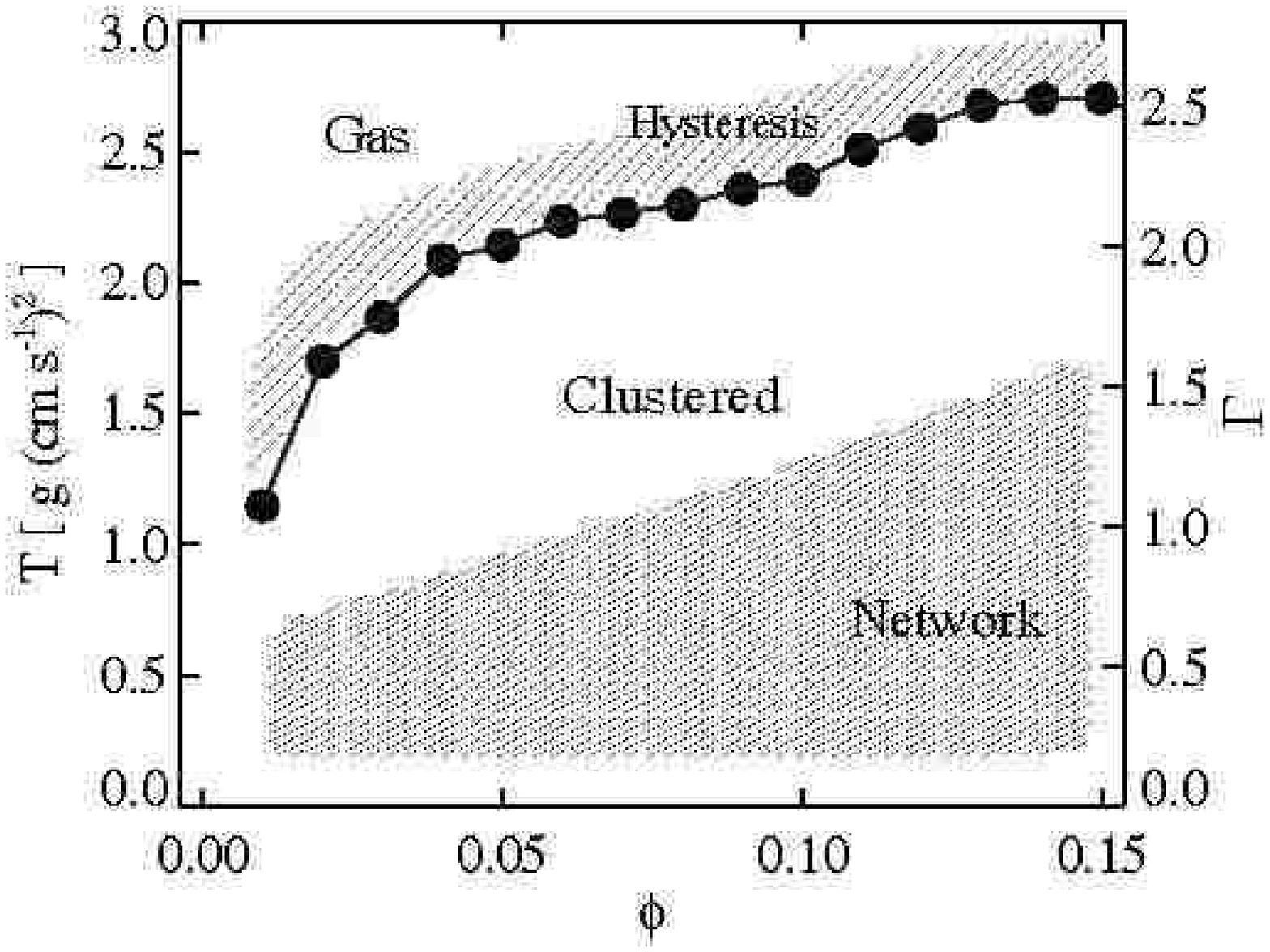}
(b)\includegraphics[width=.2\textwidth]{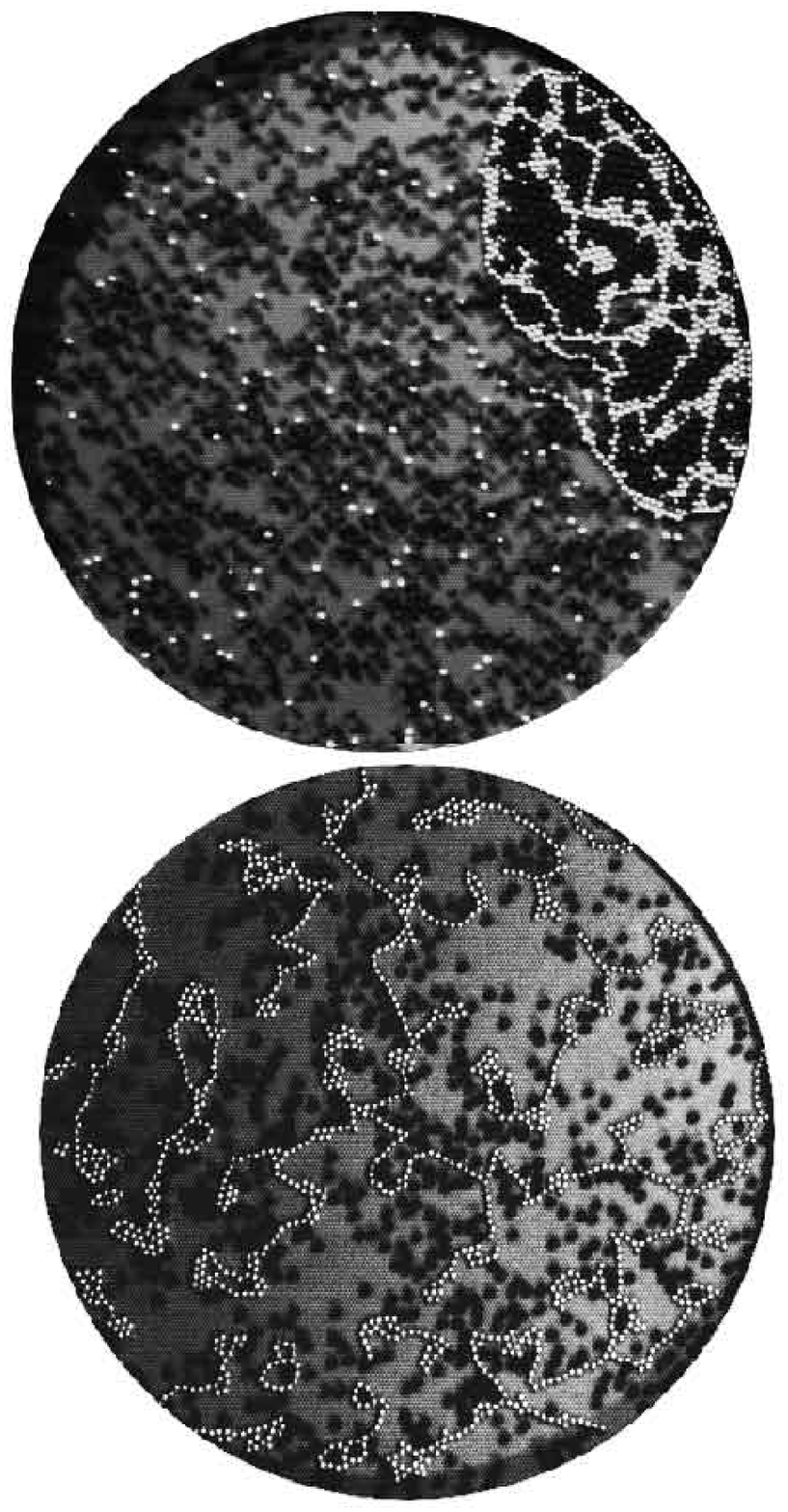}
\end{center}
\caption{(a) The magnetized particles cluster together at low vibration amplitudes, but are throughly mixed with non-magnetized glass particles at high amplitudes. The phase diagram of temperature $T$ versus the covering fraction of the particles $\phi$.  The plate acceleration $\Gamma$,
scaled by gravity, is also shown for clarity. (b) Examples of clusters formed due to magnetic interactions among non-magnetized particles. If $T$ is rapidly decreased, ramified networks of magnetized particles form. Adapted from Ref.~\cite{blair03b}.}
\label{dhs}
\end{figure}

Consider magnetic steel particles and non-magnetic glass particles in a flat bottomed container which is vibrated vertically so that energy is injected homogeneously to the particles. The particles are of the same mass but differ in size because of their different density. Consequently the granular temperature of individual particles of either kind is measured to be identical (see discussion in Section~\ref{binarytemp}). Depending on $\Gamma$, either a homogeneous gas of the two kinds of particles is found, or a clustered phase in which the magnetized particles clump together and move to the side of the container. A phase diagram and examples of the kind of clusters are shown in Fig.~\ref{dhs}. 

As $\Gamma$ is lowered, magnetized particles cluster together whereas the glass particles continue to be in a gas-like free state. Now, as the clusters form, inelasticity and friction between the particles within the cluster suppress relative motion  and particles are consequently at a lower granular temperature compared to the free particles. What is then observed is that irrespective of where a clusters forms, it will eventually end up near a side wall and stay there. 

The explanation of the migration has similar origins as the mechanism invoked to explain the experimental observation shown in Fig.~\ref{depletion}. The location of clusters near the side walls can be explained as follows. If the cluster of magnetized particles is in the center, then small particles collide with it from all sides equally. However, if the cluster is next to the side walls, the small particles collide with it from all sides except the side walls thus creating an effective force which pushes the cluster against the side walls. Thus if the temperature of the cluster is colder than the free particles, it cannot overcome this ``depletion" barrier and gets stuck against the side walls. The basic mechanism is similar to that relevant for the experiments of Aumaitre, {\em et al.} discussed in Section~\ref{vib_direc}. However, differences arise due the fact that in the experiments of Aumaitre, {\em et al.}, driving was accomplished by the side walls and therefore the cluster was found away from the side walls. A more detailed analysis of the formation and motion of the clusters can be found in Refs.~\cite{blair03b,blair_the}.\\

Detailed studies linking spatial distribution to microscopic interactions in cohesive systems have only been done recently. Indeed most of the reports discussed in this section have emerged as this review was being written.  

\section{Conclusions}

We wrap up this review by drawing some broad conclusions from the investigations that were discussed and by indicating further work that remains to be accomplished on size separation in vibrated systems. We have seen that the spatial distribution of large and small granular particles upon vibration depends on a number of factors besides size difference. The spatial distributions depend on whether the system is in a vibro-fluidized regime where particles undergo binary collisions at all times, or a regime where enduring particle-particle contact can occur. 

In the vibro-fluidized regime, it appears that the position of the large particles depends continously on the size ratio of the large and small particles and their density ratio. For high $r_d$ and low $r_\rho$, large particles may be found on top of the small particles. For $r_\rho \sim 1$, particles appear to get well mixed especially for strong vibro-fluidization. Whereas for high enough $r_\rho$, large particles will sink to the bottom. Most of the evidence for these conclusions comes from MD simulations discussed in Section~\ref{intruder_density} and some experiments~\cite{hsiau97,breu03}. Although it must be also noted that it is unclear if some of the experiments are indeed in the vibro-fluidized limit. Such experiments are feasible and need to be accomplished. A criterion for when the large particles can rise or sink has been derived using kinetic theory~\cite{jenkins02}.  However, spatial gradients in temperature and density observed in realistic vibrated granular systems have been ignored. It is vital to check the importance of such effects on the stability of the criterion. In addition, the impact of anisotropy and correlations in the particle velocity caused by their inelasticity needs to be fully analyzed.

In the regime where enduring contact between particles occurs for a substantial part of the vibration cycle, a large particle appears to always rise irrespective of density. In this regime, the friction between particles is important in supporting the weight of the large particle besides determining the energy loss during collisions. The rise of the large particle is explained in terms of a geometric void-filling mechanism where the probability of a smaller particle filling a void is greater than a large particle. Since the probabilities depend on the relative size of the particles, the rise-velocity of the large particle is observed to increase with $r_d$. 

Furthermore, the ascent velocity of the large particle (at least in the limit of a small number of large particles) appears to increase with $r_\rho$. The importance of density has been explained in terms of the relative ease with which a large particle can ``plow" through the surrounding particles as it moves upwards relative to downwards. It has been proposed that the difference arises because the granular bed is constrained at the bottom but not at the top as discussed in Section~\ref{intruder_density}. The importance of density on rise velocities indicates that a purely geometric theory will not be sufficient to explain size separation even for deep beds. 

Convection occurs frequently in deep beds due to frictional interactions with the side walls or due to the presence of interstitial air. Consequently, additional complexity can arise in the spatial patterns. As we have discussed in Section~\ref{binary_conv}, convection can destroy size separation by increasing mixing or even causing novel kinds of size separation. For example radially segregated patterns were observed when heaping occurred due to convection. 

Besides causing convection to occur, interstitial air appears to be important in determining the location of a large particle whose density is similar or less than the bed density. It has been shown that a light large particle can sink to the bottom when interstitial air is present, but will rise to the top if the air is evacuated~\cite{shinbrot98,yan03}. There is not much in the way of theory in this convection regime. In addition, few numerical simulations exist which can be used to help reduce the parameter space in order to make the development of models more feasible. 

From our discussions it is also clear that the phenomena observed in the intruder model system (limit of small number of large particles) can be different than observed with binary mixtures. One example that we discussed was where convection causes segregation of a few large particles in regions in which a broad flow converges to a thin fast flowing stream~\cite{knight93}. However, increasing the number of large particles may alter the underlying flow so that size separation may cease to occur. Another example is when a higher density intruder was observed to increase its ascent velocity whereas, segregation in binary mixtures was observed to increase when the density of the small particles was increased. The effect on spatial distributions as a function of number ratio of the two kinds of particles remains to be systematically explored. 

While we have seen that there are considerable differences between size separation in granular materials and equilibrium systems such as colloids, there are also similarities. For example, depletion forces can give rise to size separation in equilibrium systems, and a similar idea appears to be applicable to some size separation in granular systems where gravity does not play a significant role~\cite{aumaitre01}. This idea is based on the entropy-maximization principle which, at least at first, may not have seemed helpful for out of equilibrium systems. Investigations comparing and contrasting the phenomena in the two kinds of systems will not only help sharpen our understanding of each system, but may also inject new theoretical ideas into the field. 

In addition, we reviewed in Section~\ref{back} some of the studies of the basic statistical properties of not just binary mixtures but also of single species. These kinds of fundamental investigations have only recently begun and have the potential of testing and developing a theory of granular materials and size separation from the ground up. 

In this review, we have seen that a substantial body of work has been performed on the complex issue of size separation in vibrated binary granular mixtures. Although these studies have established that a particular variable (such as density, interstitial air, etc.) is important, much more work needs to be done to uncover the detailed dependence. Some of the more sophisticated experimental techniques such as magnetic resonance imaging and positron emission imaging are only recently being applied to granular flows. A detailed application of these techniques over a broad range of physical parameters still needs to be accomplished. The impact of cohesive forces on size separation needs to be investigated thoroughly. No systematic exploration of the impact of varying the dissipative properties such as inelasticity and surface friction exists, even though preliminary studies have indicated them to be important~\cite{rippie64}. 

Furthermore, some of the broad conclusions on the importance of a parameter have been drawn based from one or two direct studies. Given the sensitivity shown by granular materials to seemingly insignificant physical parameters, it would be wise to investigate the parameter space more thoroughly before putting substantial faith into our conclusions. Thus a sustained three prong approach combining experiments, simulations, and theory is still necessary to make further progress on the issue of size separation in granular materials. 

\ack

I am especially thankful to Dan Blair, Jaehyuk Choi, Mei Hou, Shu-San Hsiau, Kurt Liffman, Ingo Rehberg, Tony Rosato, and Azadeh Samadani for their help in preparing the review. This work was supported by the National Science Foundation under grant number DMR-9983659.\\


\noindent {\large \bf References} \\


\end{document}